\documentclass[epj]{svjour}   
\usepackage{graphics}  
\usepackage{epsfig}
\usepackage{bm}

\def\lessim{\lower.5ex\hbox{$\; \buildrel < \over \sim \;$}}
\def\gtrsim{\lower.5ex\hbox{$\; \buildrel > \over \sim \;$}}
\begin{document} \hbadness=10000

\title{Heavy Flavor Hadrons in Statistical Hadronization of Strangeness-rich QGP}
 \author{Inga~Kuznetsova  and Johann Rafelski}
\institute{Department of Physics, University of Arizona, Tucson, Arizona, 85721, USA}

\date{\today}

\abstract{
We study   $b$, $c$ quark hadronization from QGP.  We obtain
the yields of charm and bottom flavored hadrons within the statistical
hadronization model. The important novel feature  of this study
is that we take into
account the high strangeness and entropy content of QGP, conserving
strangeness and entropy yields at hadronization.
} 

 \maketitle
\section{Introduction}
A relatively large number of hadrons containing charmed  and bottom
quarks  are expected to be produced in heavy ion (AA) collisions at
the Large Hadrons Collider (LHC).
Because of their large mass $c,\bar c, b, \bar b$ quarks are produced
predominantly in primary
parton-parton collisions~\cite{Geiger:1993py},
at RHIC~\cite{Cacciari:2005rk}, and thus
even more so at LHC. These heavy flavor
quarks participate in the
evolution of the dense QCD matter from the beginning. In view of
the recent  RHIC results it can be
hoped that their momentum distribution could reach approximate
thermalization within the dense QGP phase~\cite{vanHees:2004gq}.

In our approach we will tacitly assume that
the following evolution stages are present in heavy-ion
collisions:
\begin{enumerate}
\item
Primary partons collide producing c, b quarks;
\item
A thermalized parton state within $\tau=\tau_{th}\simeq
0.25-1 \, \rm{fm/c}$ is formed.
By the end of this stage nearly all entropy is
produced.
\item
The subsequent chemical equilibration:
diverse thermal particle production reactions
occur, allowing first the approach to chemical equilibrium
by gluons $g$ and light non-strange quarks $q=u,d$.
\item
The strangeness chemical equilibration within $\tau\sim 5$ fm/c.
\item
The hadronization to final state near $\tau\sim 10$ fm/c.
\end{enumerate}

It is important to observe that in the presence of deconfined
QGP phase heavy hadrons containing more than one heavy quark are made from
heavy quarks created in different initial NN collisions. Therefore yields of these
hadrons are expected to be  enhanced as compared  to yields seen in single
NN collisions~\cite{Schroedter:2000ek,Becattini:2005hb}.
We note that the Bc($b\bar c,\bar b c$) and $J/\Psi (c\bar c)$
and more generally all bound
$c\bar{c}$ states yields were calculated before in the kinetic
formation and dissociation models~\cite{Schroedter:2000ek,Thews:2005fs}.
Our present work suggests that it is important to account for the binding
of heavy flavor with strangeness, an effect which depletes the eligible
supply of heavy flavor quarks which could form
Bc($b\bar c,\bar b c$) and $J/\Psi (c\bar c)$~\cite{Kuznetsova:2006hx}.

Enhanced production yield of multi-heavy hadrons can be considered to be
an indicator of the presence of deconfined QGP phase for reasons
which are analogue to those of multi-strange (anti) baryons~\cite{Koch:1986ud}.
Considering that we have little doubt that QGP is the state of matter formed in
the very high energy AA interactions, the study of yields of multi-heavy
hadrons is primarily explored in this work in order to falsify, or  justify,
features of the statistical hadronization model (SHM) employed or the model
itself in the context of formation of the heavy flavor hadrons.

For example, differing from  other recent studies   which assume    that
the hadron yields after hadronization
are in chemical equilibrium~\cite{Becattini:2005hb,Andronic:2003zv},
we form the yields based on abundance of $u,d,s$ quark pairs
as these are  available
at the chemical freeze-out (particle formation) conditions in
the quark-gluon phase. This approach is justified by the
expectation that in a fast break-up
of the QGP formed at RHIC and LHC the
phase entropy and strangeness will be nearly   conserved
during the process of hadronization.
We will investigate in quantitative terms
how such  chemical non-equilibrium yields, in the conditions we explore  well above the
chemical equilibrium abundance,  influence
the expected yields of  single, and multi-heavy flavor hadrons.

In the order
to evaluate the yields of final state hadrons we enforce
conservation of entropy, and the flavor $s,c,b$ quark pair number
during phase transition or transformation. The faster the
transition, the less likely it is that there is significant
change in strange quark pair yield. Similarly, any entropy
production is minimized when the entropy rich QGP breakup
into the entropy poor HG occurs  rapidly. The entropy conservation
constraint fixes the final light quark yield. We assume a fast
transition between QGP and HG phases, such that all hadron yields
are at the same physical conditions as in QGP breakup.

In the evaluation of heavy particle yields we form ratios involving
 as normalizer the total heavy flavor yield, and for yields
of particles with two heavy quarks we use as normalizer the product
of total yields of corresponding heavy flavors  such that the
results we consider is as little  as possible dependent on the unknown total yield
of charm and bottom at RHIC and LHC.  The order of magnitude  of the remaining
 dependence on heavy flavor yield
is set by the  ratio of yield of all particles with two heavy quarks to
yield of particles with one heavy quark. This  ratio depends  on the density
of heavy flavor at hadronization,  $(dN_c/dy)/(dV/dy)$.
The results we present for LHC are obtained for an assumed
charm and bottom quark multiplicity:
\begin{eqnarray}
 {dN_c\over dy}\equiv c&=&10  , \label{nc}\\
 {dN_b\over dy}\equiv b&=&\ 1  .  \label{nb}
\end{eqnarray}
and $dV/dy=800$ $\mathrm{fm}^3$ at $T=200$ MeV. Theoretical
cross sections of $c$ and $b$ quarks production for RHIC and LHC can
be found in~\cite{Bedjidian,Anikeev:2001rk}. In certain situations
we will explore how variation of the baseline yields Eq.\,(\ref{nc})
and  Eq.\,(\ref{nb}) impact the results. In particular among the
yields of multi-heavy hadrons, this influence  can be noticeable,
see discussion in the end of section \ref{gamcvalSec}. We note that
the number of $b$ quarks can not change during expansion, because of
large mass $m_b>>T$. It is nearly certain that all charm in QGP at
RHIC is produced in the first parton collisions, for further
discussion of LHC see Ref.\cite{Letessier:2006wn} -- it appears that
for all practical purposes also in the more extreme thermal
conditions at LHC charm is produced in the initial parton
interactions.

In order to form
physical intuition about the prevailing conditions in the QGP phase at time of
hadronization, we also evaluate the  heavy quark  chemical reference density,
that is the magnitude of the chemical occupancy factor in QGP, considering the
pre-established initial  yields of $c$ and $b$ from parton collision.
For this purpose we use  in the deconfined QGP phase:
\begin{eqnarray}
m_c &=& 1.2\ \ \mathrm{GeV},  \nonumber\\
m_b &=& 4.2.\ \ \mathrm{GeV}  \nonumber
\end{eqnarray}
We also take  ${\lambda}_i=1, i=u,d,s$ for all light flavors,
since the deviation from particle-antiparticle
yield symmetry is rather small and immaterial in the present discussion.

When computing the yields of charmed (and bottom) mesons we will distinguish
only strange and non-strange abundances, but not charged with non-charged
(e.g. D$^-(\bar c d)$ with  $\overline\mathrm{D}^0(\bar c u)$). We assume that
the experimental groups reporting results, depending on which types of D-meson
were observed, can infer the total yield (charged+non-charged) which we present.
We treat in similar way other heavy hadrons, always focusing on the heavy and
the strange  flavor content and not distinguishing the light flavor content.

Our paper is organized as follows:
we first introduce the  elements of the SHM model we use to evaluate heavy flavor hadron
yields in section \ref{SHMsec}.
This allows us to discuss the relative
yields of strange and non-strange heavy mesons in section \ref{RelCharSec},
and we show how this result relates the value of the strangeness chemical (non-)equilibrium
parameters. In this context, we also propose a multi-particle ratio as
a measure of the hadronization temperature, and explore how a multi-temperature,
staged, freeze-out would impact the relevant results.

Before proceeding to obtain the main results of this work
we introduce the notion of conservation of entropy  in section   \ref{entroSec}
and strangeness in section  \ref{strSec}, expected to be valid  in
the  fast hadronization process at LHC,
and discuss   how this impacts the SHM statistical parameters.
 We consider the entropy in
a system with evolving strangeness in subsection \ref{ssecdof} and
show that the number of active degrees of freedom in a QGP is nearly
constant. Another highlight is the discussion of sudden hadronization
of strangeness and the associated values of hadron phase space
parameters in subsection \ref{noneqSec}. Throughout this paper we will use explicitly and implicitly
the properties of QGP fireball and hadron phase space regarding entropy  and strangeness content
developed in these two sections \ref{entroSec}  and  \ref{strSec}.

We  turn to discuss the heavy flavor hadron yields
for given bulk QGP constraints in section \ref{heavyFlSec}, where we
also  compare when
appropriate to the strangeness and light quarks
chemical equilibrium results. We begin with a study of the
charm and bottom quark phase space occupancy parameters
(subsection \ref{gamcvalSec}) and turn in subsection \ref{cbMesYielSec}
to discussion of the yields of single heavy mesons, which we follow
with discussion of yields of single heavy baryons in subsection \ref{BarYieSec}.
In last subsection \ref{MultiSec} we present the expected yields of the multi-heavy hadrons, in so
far these can be considered in the grand canonical  approach.
We conclude our work with a brief summary in section \ref{concSec}.

\section{Statistical Hadronization Model (SHM)}\label{SHMsec}
The statistical hadronization model  arises from the  Fermi
multi-particle production model~\cite{Fermi:1950jd}. Fermi considered
that all hadron production matrix elements are saturated
to unity. This allows the
use of the Fermi golden rule with the N-particle phase space to
obtain the relative particle  yields. In modern language this is SHM
in micro-canonical ensemble, micro-canonical implies that
discrete (flavor) quantum numbers and the
energy are conserved exactly.

The transition from micro-canonical to
canonical, and grand-canonical ensembles  simplifies the
computational effort considerably~\cite{Hagedorn:1984hz}.
This important step does not in our context introduce
the hadron phase, although before the understanding of QGP this of
course was the reaction picture: a highly compressed hadron gas
matter evaporates particles. Today, it is
the highly compressed hot quark-gluon matter that evaporates
particles. In principle, there is no  necessity to introduce
a hadron gas phase of matter in order to use SHM
to describe particle production.

On the other hand, in order to understand
the physical meaning of the parameters introduced  to describe hadron
phase space in grand-canonical ensemble, such as temperature $T$, it is
quite convenient to {\em imagine} the existence of the hadron phase which
follows the QGP phase. This can be taken to the extreme, and a long
lasting, chemically equilibrating phase of hadrons can be assumed,
that follows in time the formation of the QGP fireball. Such a reaction
picture may not agree with the fast evolving circumstance of a heavy
ion collision. One should note that the study of hot hadron matter
on the lattice within the realm of L-QCD involves at all times
a fully equilibrated system. This will in key features differ from
the non-equilibrium QGP properties accompanying the hadro\-nization process.

The important parameters
of the SHM, which control the relative yields
of particles, are the particle specific fugacity factor ${\lambda}$ and
space occupancy factor ${\gamma}$. The fugacity is related to chemical
potential ${\mu} = T{\ln{\lambda}}$. The occupancy ${\gamma}$ is, nearly,
the ratio of produced   particles to the number of particle
expected in chemical equilibrium. The actual momentum  distribution is:
\begin{equation}
{{d^6N}\over {d^3pd^3x}}\equiv f(p)={g\over (2\pi)^3}
               {1\over \gamma^{-1}\lambda^{-1} e^{E/T}\pm 1}
             \to \gamma\lambda e^{-E/T},
\end{equation}
where the Boltzmann limit of the Fermi `$(+)$' and Bose `$(-)$' distributions is indicated,
$g$ is the degeneracy factor, $T$ is the temperature and $E=E(p)$ is the energy.

The fugacity ${\lambda}$ is associated with a
conserved quantum number, such as  net-baryon number, net-strangeness, heavy flavor.
Thus antiparticles
have inverse value of  ${\lambda}$, and  ${\lambda}$  evolution during
the reaction process is related to the
changes in   densities due to dynamics such as expansion.
 ${\gamma}$  is the same for particles and antiparticles.  Its value
changes as a function of time even if the system does not expand, for it describes buildup of
the particular particle species.
For this reason ${\gamma}$ is changes rapidly during the reaction,
while ${\lambda}$ is more constant. Thus it is ${\gamma}$ which carries
the information about the time history of the reaction and the precise
condition of particle production referred to as chemical freeze-out.

The number of particles of type `$i$' with mass $m_i$ per unit of rapidity is
in our approach given by:
\begin{equation}
\frac{dN_i}{dy}={\gamma_i}n_i^{\rm eq}\frac{dV}{dy}.  \label{dist}
\end{equation}
Here $dV/dy$ is system volume associated with the unit of rapidity,
and $n_i^{\rm eq}$ is a Boltzmann particle density in chemical equilibrium:
\begin{eqnarray}\label{BolzDis}
n_i^{\rm eq}&=&g_i\int\frac{d^3p}{(2\pi)^3}\lambda_i\exp(-\sqrt{p^2+m_i^2}/T)\nonumber\\
         &=&\lambda_i\frac{T^3}{2\pi^2}g_iW(m_i/T), \label{distapr}
\end{eqnarray}
  and
\begin{equation}\label{Wdistapr}
W(x)=x^2K_2(x)\rightarrow 2\  {\rm for}\ x\rightarrow 0.
\end{equation}
 Both, $m_ic^2\to m_i$, and $kT\to T$, are measured in
energy units when $\hbar,c,k\to 1$.

For the case of heavy flavors $m>>T$,
the dominant contribution to the Boltzmann  integral Eq.\,(\ref{BolzDis})
arises   from  $p\simeq \sqrt{2mT}$, we do
not probe the tails of the momentum distribution. Thus even when the
momentum distribution is not well thermalized, the yield of heavy flavor
hadrons  can be described in therm of the thermal yields, Eq.(\ref{dist}), where:
\begin{eqnarray}
%
%
n_i^{\rm eq}&=& \frac{g_iT^3}{2\pi^2}  \lambda_i\sqrt\frac{{\pi}m_i^3}{2T^3}{\exp}(-m_i/T)\times\nonumber\\
&&\times \left(1+\frac{15T}{8m_i}+
\frac{105}{128}\left(\frac{T}{m_i}\right)^2\ldots\right).
\label{dist1}
\end{eqnarray}
Often one can use the first term alone  for heavy flavor hadrons, since $T/m<<1$,
however the asymptotic series in Eq.\,(\ref{dist1})  has limited validity.
 Our computations are all based on CERN recursive subroutine evaluation 
of the Bessel $K_2$ functions. 

We use occupancy factors $\gamma^\mathrm{Q}_i$ and
$\gamma^\mathrm{H}_i$ for QGP and hadronic gas phase respectively, tracking
every quark flavor ($i$ = q, s, b, c) .
We assume that in the QGP phase
the light quarks and gluons are adjusting fast to the
ambient conditions, and thus are in chemical equilibrium with
$\gamma^{\mathrm{Q}}_{q,G}\to 1$.
For heavy, and strange flavor, the value of
$\gamma^{\mathrm{Q}}_i$ at hadronization condition is  given  by the number of
particles present, generated by prior kinetic processes, see Eq.\,(\ref{dist}).

The  yields of different quark flavors
originate in different physical processes, such as production in
initial collisions for $c,b,s$, and for $s$ also production in
thermal plasma processes.
In general we thus cannot expect  that   $\gamma^{\mathrm{Q}}_{c,b}$ will be near unity
at hadronization. However, the thermal strangeness production process  $GG\to s\bar s$
can nearly chemically equilibrate strangeness flavor in plasma
formed at RHIC and/or LHC~\cite{Letessier:2006wn},
and we will always consider, among other cases the limit
$\gamma^{\mathrm{Q}}_s\to 1$ prior to
hadronization.

The yields of all
hadrons after hadronization are also given by Eq.\,(\ref{dist}),
which helps us to obtain $\gamma^H_{i}$  for hadrons. In general,
the evaluation of hadron chemical parameters presents a  more
complicated case since they are composed from those of valance quarks
in the hadron (three quarks, or a quark and anti quark).
Therefore, in the coalescence picture, the phase space occupancy
$\gamma^\mathrm{H}$ of hadrons will be the product of
$\gamma^\mathrm{H}$'s for each constituent quark. For example for
the charmed meson D ($c\bar{q}$)
\begin{equation}
\gamma^\mathrm{H}_D = \gamma^\mathrm{H}_c\gamma^\mathrm{H}_q.
\end{equation}

To evaluate yields of final state hadrons we enforce
conservation of entropy, and the flavor $s,c,b$ quark pair number
during phase transition or transformation. The faster the
transition, the less likely is that there is significant
change in strange quark pair yield. Similarly, any entropy
production is minimized when the entropy rich QGP breakup
into the entropy poor HG occurs  rapidly. The entropy conservation
constraint fixes the final light quark yield. We assume a fast
transition between QGP and HG phases, such that all hadron yields
are at the same physical conditions as in QGP breakup.

Assuming that in the hadronization process the number of
$b$, $c$, $s$ quark pairs doesn't change,
the three unknown $\gamma^\mathrm{H}_s$,
$\gamma^\mathrm{H}_c$,
$\gamma^\mathrm{H}_b$ can be determine from their
values in the QGP phase, $\gamma^\mathrm{Q}_s$,
$\gamma^\mathrm{Q}_c$,
$\gamma^\mathrm{Q}_b$ (or $dN^Q_i/dy$) and the three flavor conservation
equations,
\begin{equation} \label{flcons}
\frac{dN^\mathrm{H}_i}{dy}=\frac{dN^\mathrm{Q}_i}{dy}=\frac{dN_i}{dy}, \quad i=s,c,b.
\end{equation}
In order to conserve entropy:
\begin{equation}
\frac{dS^\mathrm{H}}{dy}=\frac{dS^\mathrm{Q} }{dy}=\frac{dS}{dy}, \label{Scons}
\end{equation}
a value $\gamma^\mathrm{H}_q\ne 1$ is nearly always required when in
the QGP phase $\gamma^\mathrm{Q}_{q,G} = 1$.
This implies that yields of hadrons with light quark content
are, in general, not in chemical equilibrium, unless there is some extraordinary circumstance
allowing a prolonged period of time  in which hadron reactions can occur after hadronization. Chemical
non-equilibrium thus will influence the yields of  heavy
flavored particles in final state as we shall discuss in this work.

As noted at the beginning of this section,
the use of the hadron phase space (denoted by H above)
does not imply the presence of a real physical `hadron matter' phase: the  SHM
particle yields will be attained  solely on the basis of availability of this  phase
space as noted at the beginning of this section.  Another way to argue this is to
imagine a pot of quark matter with hadrons evaporating . Which kind
of hadron emerges and at which momentum is entirely determined by the access to
the phase space, and there are only free-streaming particles in the final state.

Thinking in these terms, one can imagine that especially
for heavy quark hadrons some particles are
pre-formed in the deconfined plasma, and thus the heavy hadron yields
may be based on a value of temperature which is higher than the global
value expected  for other hadrons.
For this reason we will study in this work  a range $140<T<260$ MeV
and also consider sensitivity to this type of two-temperature chemical
freeze-out of certain heavy hadron yield ratios.

\section{Relative charmed hadron yields}\label{RelCharSec}
\subsection{Determination of  $\gamma_s/\gamma_q$}\label{RelGamSec}
We have seen  considering $s/S$ and also $s$ and $S$ individually
across the phase limit that in general one would expect
chemical non-equilibrium in hadronization
of chemically equilibrated QGP.
We first show that this result
matters for the relative charm meson yield ratio $D_s/D$, where
$D_s(c\bar{s})$ comprises all mesons of type $(c\bar{s})$ which
 are   listed in the bottom section of table \ref{openbc}, and
 $D(c\bar{q})$ comprise yields of all $(c\bar{q})$states  listed in the top
section of table~\ref{openbc}. This ratio is formed
based on the assumption that on the time scale of strong interactions
the family of strange-charmed mesons can  be distinguished
from the family non-strange charmed mesons.

\begin{table}
\caption{Open charm, and bottom, hadron states we considered. States
in parenthesis either need confirmation or have not been observed experimentally,
in which case  we follow the values of Refs.\,\cite{Cheu:2004zc,Matsuki:1997da}.
We implemented charm-bottom symmetry required for certain observables.
Top section: $D$, $B$ mesons, bottom section: $D_s$, $B_s$-mesons
\label{openbc}}
\begin{tabular}{|c|c|c|c|c|c|}
  \hline
   hadron& M[GeV] &Q:c,b&hadron&  M[GeV]&g\\
  \hline 
$D^0(0^-)$& 1.8646&$Q\bar{u}$&$B^0(0^-)$&5.279&1\\
$D^+(0^-)$ & 1.8694&$Q\bar{d}$&$B^+(0^-)$&5.279&1\\
$D^{*0}(1^-)$&2.0067&$Q\bar{u}$&$B^{*0}(1^-)$&5.325&3\\
$D^{*+}(1^-)$&2.0100&$Q\bar{d}$&$B^{*+}(1^-)$&5.325&3\\
$D^0(0^+)$ &2.352&$Q\bar{u}$&$B^0(0^+)$&5.697&1\\
$D^+(0^+)$&2.403&$Q\bar{d}$&$B^+(0^+)$&5.697&1\\
$D^{*0}_1(1^+)$&2.4222&$Q\bar{u}$&$B_1^{*0}(1^+)$&5.720&3\\
$D^{*+}_1(1^+)$&2.4222&$Q\bar{d}$&$B_1^{*+}(1^+)$&5.720&3\\
$D^{*0}_2(2^+)$&2.4589&$Q\bar{u}$&$B^{*0}_2(2^-)$&(5.730)&5\\
 $D^{*+}_2(2^+)$&2.4590&$Q\bar{d}$&$B^{*+}_2(2^+)$&(5.730)&5\\
  \hline  
 $D_s^+(0^-)$&1.9868&$Q\bar{s}$&${B_s^0}(0^-)$&5.3696&1\\
 $D^{*+}_s(1^-)$&2.112&$Q\bar{s}$&${B^{*0}_s}(1^-)$&5.416&3\\
 $D^{*+}_{sJ}(0^+)$&2.317&$Q\bar{s}$&${B^{*0}_{sJ}}(0^+)$&(5.716)&1\\
 $D^{*+}_{sJ}(1^+)$&2.4593&$Q\bar{s}$&${B^{*0}_{sJ}}(1^+)$&(5.760)&3\\
 $D^{*+}_{sJ}(2^+)$&2.573&$Q\bar{s}$&${B^{*0}_{sJ}}(2^+)$&(5.850)&5\\
  \hline
\end{tabular}
\end{table}

\begin{figure}
\centering
\includegraphics[width=7.8cm,height=7.8cm]{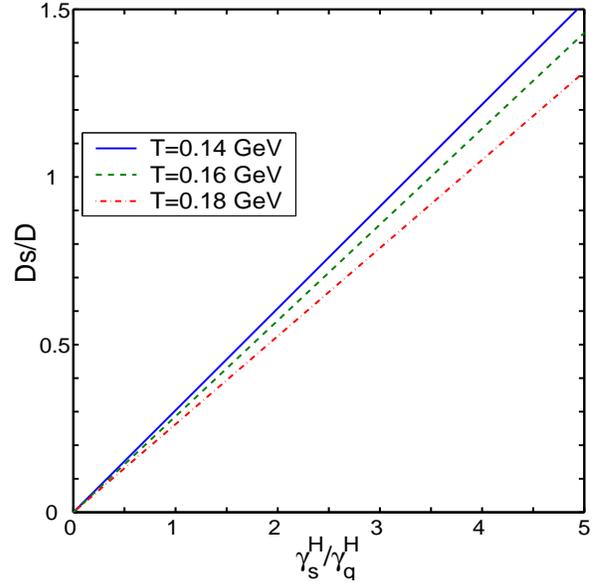}
\caption{(Color on line) \small{$D/D_s$ ratio as a function of
$\gamma^\mathrm{H}_s/\gamma^\mathrm{H}_q$ for $T = 140$ MeV (blue,
solid line), $T = 160$ MeV (green, dashed line) and $T= 180$ MeV
(red, dash-dot line)}. } \label{rDsDg}
\end{figure}

\begin{figure}[!hbt]
\centering
\includegraphics[width=8cm,height=8cm]{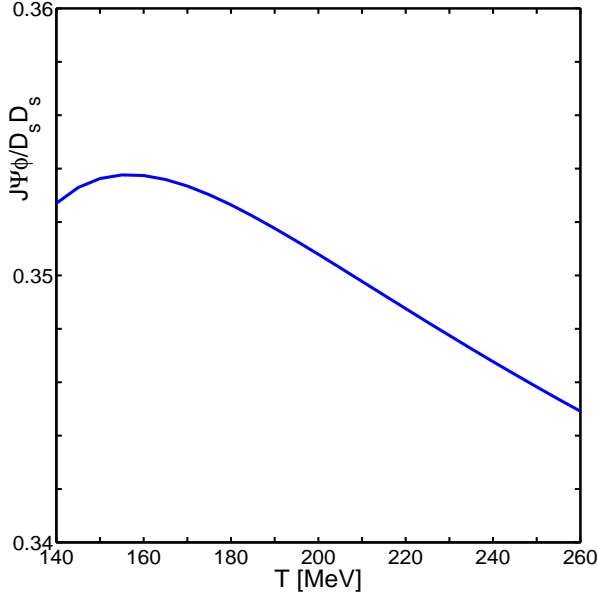}
\caption{(Color on line)
\small{$J\!/\!\Psi\,\phi/D_s{\overline{D_s}}$ ratio  as a function of hadronization temperature
T.}} \label{JpsiD}
\end{figure}

The yield ratio $D_s/D$ calculated using Eq.\,(\ref{dist}) and
Eq.\,(\ref{distapr}) is shown in figure~\ref{rDsDg}. Using
Eq.\,(\ref{dist1})  we see that this ratio is proportional to
${\gamma^{\rm H}_s}/{\gamma^{\rm H}_q}$  and weakly dependent on
$T$:
\begin{equation}
\frac{D_s}{D}\approx\frac{\gamma^{\rm H}_s}{\gamma^{\rm{H}}_q}\frac{\Sigma_i
g_{Dsi}m_{Dsi}^{3/2}\exp(-m_{Dsi}/T)}{\Sigma_i
g_{Di}m_{Di}^{3/2}\exp(-m_{Di}/T)}=f(T)\frac{\gamma^{\rm H}_s}{\gamma^{\rm H}_q}.
\end{equation}
A deviation of
${\gamma_s}/{\gamma_q}$
from unity in the range we will see in section \ref{noneqSec}
leads to a noticeable difference in the ratio $D_s/D$.
We show in figure~\ref{rDsDg} results for $T=140, 160, 180$ MeV.
In this $T$-domain, the effect due to
${\gamma^{\rm H}_s}/{\gamma^{rm H}_q}\ne 1$
is the dominant contribution to the variation of this relative yield.

\subsection{Check of statistical hadronization model}\label{singHadSec}
We next   construct a heavy flavor particle ratio that depends on
hadronization temperature  only. To cancel the fugacities and  the
volume we consider the ratio $J\!/\!{\Psi}\phi/D_s\overline{D_s}$ in
figure~{\ref{JpsiD}}. Here $J\!/\!\Psi$ yield  includes the  yield
of $(c\bar c)$ mesons decaying into the $J\!/\!\Psi$.  All phase
space occupancies cancel since $J/\Psi \propto
\gamma^{\mathrm{H}\,2}_c$, $\phi \propto \gamma^{\rm H\,2}_s$, $D_s
\propto \gamma^{\mathrm{H}}_c\gamma^{\rm H}_s$ and similarly
$\overline{Ds}\propto \gamma^{\mathrm{H}}_c\gamma^{\rm H}_s$. When
using here the particle $D_s(\bar c s)$ and antiparticle
$\overline{D_s}(c \bar s)$    any chemical potentials present are
canceled as well. However, for the LHC and even RHIC environments
this refinement is immaterial.

\begin{figure}
\centering
\includegraphics[width=9cm,height=9cm]{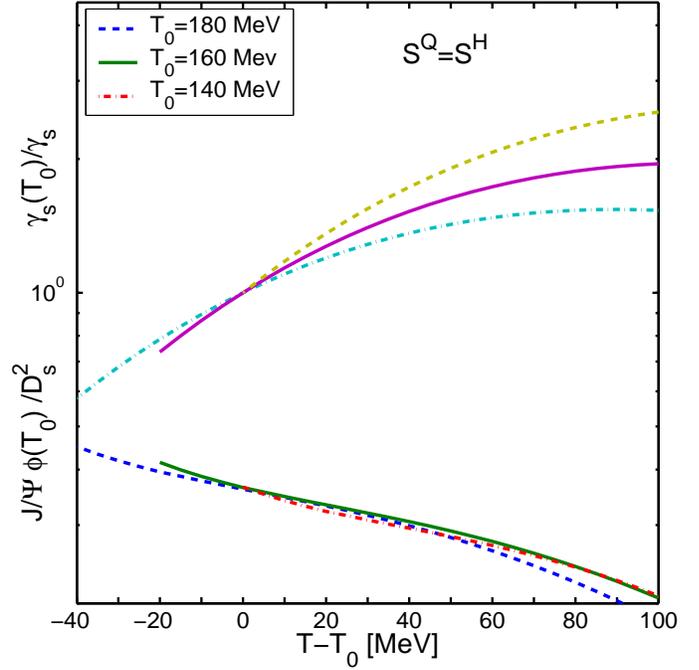}
\caption{(Color on line) \small{$J\!/\!\Psi\,\phi(T_0)/D_sD_s$ ratio
is evaluated at two temperatures, $T $ for  heavy flavor hadrons,
and $T_0$ for  $\phi$ as a function of $T-T_0$, with three values of
$T_0=140,160,180$ MeV is considered  with $S^H=S^Q$. }}
\label{JpsiD2T}
\end{figure}

This ratio  $J\!/\!{\Psi}\phi/D_s\overline{D_s}$, turns out to be
practically constant, within a rather wide  range of  hadronization
temperature $T$, see figure~{\ref{JpsiD}}.
The  temperature range we study $140<T<280$ MeV allows us to consider
an early freeze-out of different hadrons. To be sure
of the temperature independence  of $J\!/\!{\Psi}\phi/D_s\overline{D_s}$
we next consider the possibility that
hadronization temperature $T$ of charmed hadrons is higher than
hadronization temperature $T_0$ of $\phi$. We study this
question by exploring the sensitivity of the ratio
 $J\!/\!{\Psi}\phi/D_s\overline{D_s}$ to the two temperature
freeze-out in figure \ref{JpsiD2T}, see bottom three
lines for $T_0=180,160,140$ MeV  with $\gamma_q$ from condition $S^Q=S^H$, see figure~\ref{gq}.  If charmed hadrons hadronize later, $T>T_0$, and $T-T_0<60$
the change in $J\!/\!{\Psi}\phi/D_s\overline{D_s}$ ratio is small (about 20\%).
If this were to be
measured as experimental result,
\begin{equation}
{J\!/\!{\Psi}\phi\over D_s\overline{D_s}}\simeq 0.35,
\end{equation}
one could not but conclude that
all particles involved are formed by mechanism of statistical hadronization.

This interesting result can be understood, considering the behavior
of the $\gamma^{\rm H}_s(T_0)/\gamma^{\rm H}_s(T)$ ratio, which
increases rapidly  with increasing $T-T_0$ (see the top three lines
in figure \ref{JpsiD2T}). This ratio almost compensates the change
in $\phi$-yield, an effect we already encountered in the context of
the results we show below in figure \ref{phiK}. For large
$T-T_0$ the ratio ${J\!/\!{\Psi}\phi/ D_s\overline{D_s}}$ begins to
decrease more rapidly because $\gamma_s$ increases for $S^H=S^Q$,
see figure~(\ref{gq}).

\section{Entropy in Hadronization}\label{entroSec}
\subsection{Entropy in QGP fireball}\label{sseentr}
The  entropy content is seen in the  final state multiplicity of
particles produced after hadronization. More specifically, there is a relation
between entropy and particles multiplicities, once we note that the
entropy per particle in a gas is:
\begin{equation}
\frac{S_{\mathrm {B}}}{N}=3.61,\ \ \ \ \ \frac{S_\mathrm{cl}}{N}=4,\
\ \ \ \ \frac{S_\mathrm{F}}{N}=4.2,
\end{equation}
for massless Bose, classical (Boltzmann) and Fermi gases, respectively.
Effectively, for QGP with $u,s,d,G$ degrees of freedom,
$S^{\mathrm {Q}}/N^{\mathrm {Q}}\sim 4$ is applicable for large
range of masses. Thus:
\begin{equation}
\frac{dS^{\mathrm{Q}}}{dy} \approx 4\,\frac{dN^Q}{dy}. \label{ent}
\end{equation}
This in turn means that final state particle multiplicity provides
us with information about the primary entropy content generated in the
initial state of the QGP phase.

It is today generally believed that there is entropy conserving hydrodynamic
expansion of the QGP liquid. Entropy is conserved in the
fireball, and the conservation of entropy density $\sigma$ flow is expressed by:
\begin{equation}
\frac{\partial_\mu(\sigma u^{\mu})}{\partial{x^{\mu}}}=0,
\label{flowcon} \label{entrfl}
\end{equation}
where $u^{\mu}$ is local four  velocity vector.  A special case of
interest is   the so-called Bj{\o}rken scenario~\cite{Bjorken:1982qr}
for which Eq.\,(\ref{entrfl})  can be solved exactly assuming as
initial condition scaling of the physical properties
as a function of rapidity. This implies that there is no preferred
frame of reference, a situation expected in very high energy collisions.
Even if highly idealized, this  simple reaction picture allows
a good estimate of many physical features. Of relevance here is that the exact
solution of hydrodynamics in  (1+1) dimensions implies
\begin{equation}
\frac{dS}{dy}=\rm{Const}.\label{entcon1}\,.
\end{equation}

Thus entropy $S$ is not only conserved globally in the hydrodynamic
expansion, but also per unit of rapidity. Though we have (1+3) expansion,
Eq.({\ref{entcon1}}) holds as long as there is, in rapidity, a
flat plateau of particles yields. Namely, each of the
domains of rapidity is equivalent,  excluding  the projectile-target
domains. However, at RHIC and LHC energies  these
are causally disconnected from the central rapidity bin, where
we study the evolution of heavy flavor. The entropy we observe in
the final hadron state has been  to a large extent   produced after the heavy
flavor is produced, during the initial
parton thermalization phase, but before strangeness has been produced.
In order to model production of hadrons for different chemical freeze-out
scenarios of the same reaction, we need to relate the entropy content,
temperature and volume of the QGP domain.   We  consider for a
$u,d,G$-chemically equilibrated QGP, and allowing for partial chemical
equilibration of strangeness, the entropy content.

The entropy density $\sigma$ can be obtained from the equation
\begin{equation}\label{sigdef}
\sigma\equiv\frac{S}{V}=-\frac{1}{V}\frac{d{F_\mathrm{Q}}}{dT}, \label{entden}
\end{equation}
where the thermodynamic potential is:
\begin{equation}
F_\mathrm{Q}(T, \lambda_q, V)=-T\ln {Z(_\mathrm{Q}T, \lambda_q, V)}_\mathrm{Q}.
\label{thpot}
\end{equation}
Inside QGP the partition function is a product of partition
function of gluons $Z_g$, light quarks $Z_q$ and strange quarks  $Z_s$, hence:
\begin{equation}
\ln{Z}=\ln{Z_g}+\ln{Z_q}+\ln{Z_s};
\end{equation}
where for massless particles with $\lambda_q=1$
\begin{eqnarray}
\ln{Z_g}&=&\frac{g_g\pi^2}{90}VT^3,\\
\ln{Z_q}&=&\frac{7}{4}\frac{g_q\pi^2}{90}VT^3.
\end{eqnarray}
Here $g_g$ is degeneracy factor for gluons and $g_q$ is degeneracy factor
for quarks. The factor $7/4=2\cdot 7/8$ accounts for the difference in statistics and
presence of both quarks and antiquarks. The number of degrees of freedom of
quarks and gluons is influenced by  strongly  interactions, characterized
by strong coupling constant $\alpha_\mathrm{s}$:
\begin{eqnarray}
g_g &=& 2_s\,8_c\,\left(1-\frac{15}{4\pi}{\alpha_\mathrm{s}}+\ldots\right); \\
g_q&=&2_s\,3_c\,2_f\,\left(1-\frac{50}{21\pi}{\alpha_\mathrm{s}}+\ldots\right).
\end{eqnarray}

The case of strange quarks is somewhat more complicated, since we have
to consider the mass, the degree of chemical equilibration, and guess-estimate
the strength of QCD perturbative interactions. We have in Boltzman approximation:
\begin{eqnarray}
\ln{Z_s}&=&2_{\rm p/a}\frac{g_s}{\pi^2} VT^3,\\
\label{gsk}
g_s &=& 2_s3_c\gamma_s^{\rm Q} \,0.5W(m_s/T)\left(1-k\frac{\alpha_s}{\pi}\right).
\end{eqnarray}
$W(m/T)$ is function seen in Eq.\,(\ref{Wdistapr}).
We allow both for strange and antistrange quarks,
factor $2_{\rm p/a}$ (which is for massless fermions $2\cdot 7/8=7/4$).
$k$ at this point is a temperature dependent parameter.
Even in the lowest order perturbation theory it
has not been evaluated for massive quarks at finite
temperature. We know that for massless quarks  $k\simeq 2$.
Considering expansion in $m/T$, for large masses
the correction reverses sign~\cite{Kapusta:1979fh},
which result supports the reduction in value of $k$
for $m\simeq T$. We will
use here the value $k=1$~\cite{Letessier:2006wn}.

\subsection{Number of degrees of freedom in QGP}\label{ssecdof}
The entropy density  following from Eq.\,(\ref{sigdef}) is:
\begin{equation}\label{entrcons}
\sigma=\frac{4\pi^2}{90}(g_g + \frac{7}{4}g_q)T^3
      +\frac{4}{\pi^2}2_{\rm p/a}g_sT^3
      +\frac{\cal{A}}{T}.
\end{equation}
For strange quarks in the  second term in Eq.\,(\ref{entrcons})
we set the entropy per strange quarks to 4 units.
In choosing $S_s/N_s=4$ irrespective of the effect of interaction
and mass value $m_s/T$ we are minimizing the influence of  unknown
QCD interaction effect.

The last term  in Eq.\,(\ref{entrcons}) comes from differentiation of the
strong coupling constant $\alpha_\mathrm{s}$ in the partition function
with respect to  $T$, see Eq.({\ref{entden}}). Up to two loops in
the $\beta$-function of the renormalization group the correction
term is~\cite{Hamieh:2000fh}:
\begin{equation}
{\cal{A}}=(b_0{\alpha^2_s}+b_1{\alpha^3_s})\left[\frac{2\pi}{3}T^4+\frac{n_f5\pi}{18}T^4\right]
\end{equation}
with $n_f$ being the number of active fermions in the quark loop, $n_f\simeq 2.5$, and
\begin{eqnarray}
b_0 = \frac{1}{2\pi}\left(11-\frac{2}{3}n_f\right),\
b_1 = \frac{1}{4\pi^2}\left(51-\frac{19}{3}n_f\right).
\end{eqnarray}
For the strong coupling constant $\alpha_\mathrm{s}$ we use
\begin{equation}
\alpha_s(T) \simeq \frac{\alpha_s(T_c)}{1+{C}\ln(T/T_c)}, \ \ \
C=0.760\pm 0.002,
\end{equation}
where $T_{c}=0.16\,$GeV. This expression arises from renormalization group running
of $\alpha_s(\mu)$, the energy scale at $\mu=2\pi T$, and the value $\alpha_s(M_Z)=0.118$.
A much more sophisticated study
of the entropy in the QGP phase is possible~\cite{Kapusta}, what we use
here is an effective model which agrees  with the
lattice data~\cite{Letessier:2003uj}.

Eq.\,(\ref{entrcons}) suggests that we  introduce an effective degeneracy of the QGP
based on the expression we use for entropy:
\begin{equation}\label{gS}
g_{\rm eff}^Q(T) = g_g(T) + \frac{7}{4}g_q(T)
+ 2g_s\frac{90}{\pi^4}+\frac{\cal{A}}{T^4}\frac{90}{4\pi^2}.
\end{equation}
Which allows us to write:
\begin{equation}\label{entrcons1}
\sigma=\frac{4\pi^2}{90}g_{\rm eff}^QT^3,
\end{equation}
and
\begin{equation}\label{SdVdy}
\frac{dS}{dy}=\frac{4\pi^2}{90}g_{\rm eff}^QT^3\frac{dV}{dy}\simeq\rm{Const}.
\end{equation}

\begin{figure}
\centering
\includegraphics[width=8cm]{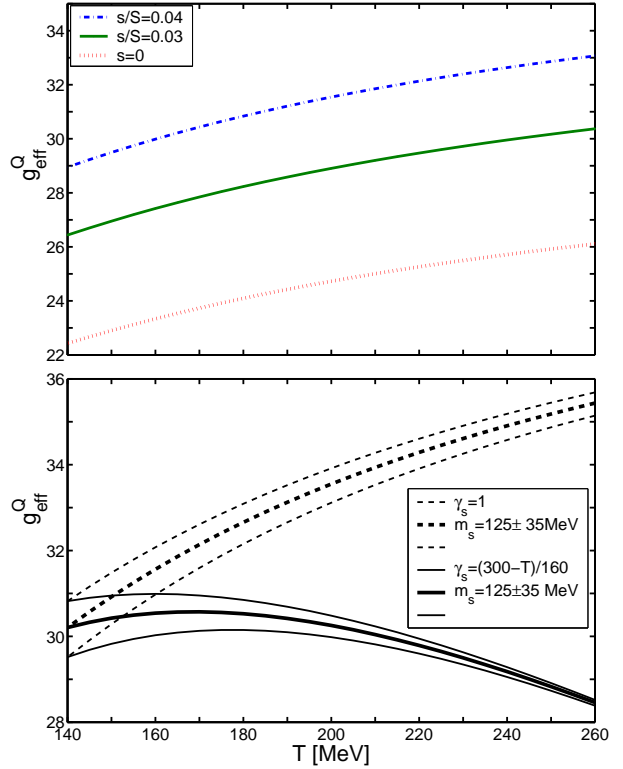}
\caption{(color on-line)
\small{
The Stefan-Boltzman degrees of freedom $g_{\rm eff}$   based on
entropy content of QGP, as function of temperature $T$.
Upper frame:   fixed $s/S$, the solid line with dots (green)
is for a system  with fixed strangeness per entropy $s/S=0.03$,
while dot-dashed (blue) line is for  $s/S=0.04$. The dotted (red)
line is for 2-flavor QCD $s/S=0$ ($u,d,G$ only);
The bottom frame shows dashed (black) line
2+1-flavor QCD with  $m_s=125\pm35$\,MeV (chemically equilibrated
$u,d,s,G$ system). The (thick, thin) solid lines are
for QGP in which strangeness contents is increasing as a function of
temperature, see text. }} \label{geff}
\end{figure}

We show the QGP degeneracy  in figure \ref{geff},  as a function
of $T\in [140,260]$ MeV,  top frame for
fixed $s/S=0, 0.03, 0.04$ (from bottom to top), and in the bottom
frame for the strangeness chemical  equilibrium, $\gamma_s=1$ (dashed)
and approach to chemical equilibrium cases (solid).
When we fix the specific strangeness content $s/S$  in the plasma
comparing different temperatures we find  that in all cases
$g_{\rm eff}^Q$  increases with $T$.
For $s/S=0$ we have a 2-flavor system (dotted line, red) and
the effective number of degrees of freedom $g_{\rm eff}^Q$
varies between 22 and 26.   The
solid line with dots (green) is for $s/S=0.03$, and the dot-dashed line (blue)
gives the result for  $s/S=0.04$.

In the bottom panel of figure \ref{geff} we note that like for 2 flavors,
case (s=0), for the 2+1-flavor system ($\gamma_s=1$) $g_{\rm eff}^Q$ increases
with T (dashed line, black). $g_{\rm eff}^Q$ varies between 30 and 35.5.
The thin dashed lines indicate the range of uncertainty due to mass of the
strange quark, which in this calculation is fixed with upper curve
corresponding to $m_s=90$ MeV, and lower one $m_s=160$ MeV. The expected
decrease in value of $m_s$ with $T$ will thus have the effect to steepen
the rise in the degrees of freedom with $T$.

We now explore in a QGP phase  the effect of an increasing
strangeness fugacity with decreasing temperature. This study is
a bit different from the rest of this paper, where we consider
for comparison purposes hadronization for a range of temperatures
but at a {\em fixed} value of $s/S$. A variable $\gamma_s^{\rm Q}(T)$
implies a more sophisticated, and thus more model dependent
picture of plasma evolution. However, this offers us an important insight
about  $g_{\rm eff}^Q$.

We consider the function:
\begin{equation}\label{gamsmodel}
\gamma_s^{\rm Q}={300-T{\rm [MeV]} \over 160}.
\end{equation}
This is  consistent with the kinetic computation of
strange\-ness production~\cite{Letessier:2006wn}.
 At $T=140$ MeV  we have chemical equilibrium in the QGP phase,
while and at the  temperature $T=260$\,MeV  we  have
$\gamma_s=0.25$.
The result for   $g_{\rm eff}^Q$ is shown as a thick (black) line in figure \ref{geff},
with the range showing   strange quark mass range $m_s=125\pm35$\,MeV.
 We see that in a wide range of temperatures
we have $29.5<g_{\rm eff}^Q<30.5$.

The lesson is that  with
the growth of $\gamma_s^{\rm Q}$ with decreasing $T$
the entropy of the QGP is well described by
a constant value  $g_{\rm eff}^Q=30\pm0.5$. Since the entropy is (nearly) conserved
and $g_{\rm eff}^Q$ is (nearly) constant, Eq.\,(\ref{SdVdy}) implies that
 we can scale the system properties using the constraint $T^3dV/dy=$Const.
We stress again that these results  arises
in a realistic QGP with $2+\gamma_s^{\rm Q}$-flavors, but are model dependent
and of course rely on the lattice motivated description of
the  behavior of QGP properties.  On the other hand it is not surprising that
the rise of
strangeness chemical saturation with decreasing temperature compensates   the
`freezing'  of the  $q,G$-degrees of freedom
with decreasing temperature.

\subsection{Entropy content and chemical (non-)equilibrium}\label{ssentrocont}

We use as a reference a QGP state with
$dV/dy=800\,{\mathrm{fm}}^{3}$ at $T=200\,{\mathrm{MeV}}$,
see table \ref{VTN}. We find  from Eq.\,(\ref{ent})
the Q and H phase particle multiplicity. The hadron multiplicity stated
is what results after secondary resonance decays. The  total hadron
multiplicity after hadronization and resonance decays
was calculated using on-line SHARE 2.1~\cite{Torrieri:2004zz}.
If a greater (smaller) yield of final state hadrons is observed at LHC,
the value of $dV/dy$ need to be revised up (down). In general
expansion before hadronization
will not alter $dS/dy$. We can  expect that as $T$ decreases,
$V^{1/3}$ increases. Stretching the validity of Eq.\,(\ref{SdVdy}) to
low temperature  $T=140$ MeV, we see the result in the second line of table \ref{VTN}.

\begin{table}
\caption{Reference values of  volume, temperature, entropy, particle multiplicity}
 \label{VTN}
\begin{tabular}{|c|c|c|c|c|}
  \hline
 $dV/dy$ $[\mathrm fm^{3}]$ & $T$[MeV]  & $dS^{\mathrm {Q}}/dy$ & $dN^{\mathrm {Q}}/dy $& $dN^{\mathrm {H}}/dy$\\
  \hline
  $800$& $200$ &10,970&2,700&5,000\\
  $2300$& $140$ &10,890&2,700&4,500\\
\hline
\end{tabular}
\end{table}

For QGP, in general  the entropy content is higher than
in a comparable volume of chemically equilibrated
hadron matter,  because of the liberation of color degrees of freedom in the
color-deconfined phase. The total entropy has to be
conserved during transition between QGP and HG phases, and
thus after hadronization, the excess of entropy is
observed in excess particle multiplicity, which can be interpreted as a
signature of deconfinement~\cite{Letessier:1992xd,Letessier:1993hi}.
The dynamics of the transformation of QGP into HG determines how this
additional entropy manifests itself.

The comparison of entropy in both phases is temperature
dependent but in the domain of interest i.e.
$140<T<180$ MeV  the entropy density follows:
\begin{equation}
\sigma^{\mathrm{Q}} \gtrsim 3 \sigma^{\mathrm{H}}.
\end{equation}
Since the total entropy $S$ is conserved or slightly increases,
in the hadronization process some key parameter must grow in the
hadronization process. There are two options:\\
a) either the volume changes:
\begin{equation}
3 V^{\mathrm{H}} \gtrsim V^{\mathrm{Q}};
\label{volch}
\end{equation}
or \\
b) the phase occupancies change, and since $n_i\propto \gamma_i^{2,3}, i=q,s$ in
hadron phase
\begin{equation}
\gamma^{\mathrm{H}}_q\simeq \sqrt{3}, \ \ \ \gamma^{\mathrm{H}}_s/\gamma^{\mathrm{H}}_q\gtrsim 1.
\end{equation}
In a slow, on hadronic time scale, transition, such as is the case in the
early Universe, we can expect that case a) prevails. In
high energy  heavy ion collisions, there is
no evidence in the experimental results   for the long coexistence of hadron and
quark phases which is required for volume growth. Consequently, we have
$V^{\mathrm{H}}\sim V^{\mathrm{Q}}$ and
a large value of $\gamma^{\mathrm{H}}_q$ is required
to conserve entropy. The value of
$\gamma^\mathrm{H}_q$ is restricted by
\begin{equation}
\gamma^\mathrm{cr}_q\cong\exp(m_{\pi}^0/2T). \label{bcon}
\end{equation}
This value $\gamma^\mathrm{cr}_q$  is near to maximum
allowed value, which arises at condition of Bose-Einstein
condensation of pions. We will discuss quantitative results for
$\gamma^{\mathrm{H}}_q$ (and $\gamma^{\mathrm{H}}_s$) below
in subsection  \ref{noneqSec}.

\section{Strangeness in Hadronization}\label{strSec}
\subsection{Abundance in QGP and HG}\label{StrAbund}
The efficiency of strangeness production depends on energy and collision
centrality of heavy ions collisions. The increase, with value of
centrality (participant number), of
per-baryon specific strangeness yield indicates presents of
strangeness production mechanism acting beyond the first collision
dynamics. The thermal gluon fusion to strangeness can explain this
behavior~\cite{Koch:1986ud},
and a model of the flow dynamics at RHIC and LHC suggests
that the QGP approaches chemical equilibrium but also can
exceed it at time of hadronization~\cite{Letessier:2006wn}.

The strangeness yield in  chemically equilibrated  QGP is usually
described as an ideal Boltzman gas. However, a significant correction
is expected due to perturbative QCD effects. We implement this
correction based on   comments below Eq.\,(\ref{gsk}). We use here the expression:
\begin{equation}
\frac{dN_{s}^{\rm Q}}{dy}=\gamma_s^Q\left(1-\frac{\alpha_s}{\pi}\right)n^{\rm eq}_s\frac{dV}{dy}.\label{str}
\end{equation}
with the Boltzman limit density Eq.\,(\ref{BolzDis}), and mass $m_s=125$\,MeV,
$g_s=6, \lambda_s=1$. The QCD correction corresponds to discussion
of entropy in subsection \ref{sseentr}

We obtain strange quarks phase space occupancy
$\gamma^{\mathrm{H}}_s$ as a function of temperature from
condition of equality of the number of strange quark and antiquark
pairs in QGP and HG. Specifically,
in the sudden QGP hadronization,  quarks recombine and we expect that the
strangeness content does not significantly change.
For heavier flavors across the phase boundary
this condition  Eq.\,(\ref {flcons}) is very well satisfied,  for strangeness the
fragmentation effect adds somewhat to the yield,
\begin{equation}\label{sconshad}
 \frac{dN_{s}^{\rm H}}{dy} \gtrsim \frac{dN_{s}^{\rm Q}}{dy}.
\end{equation}
Using the equality of yields we underestimate slightly the value of strangeness
occupancy that results. We recall that we also conserver entropy  Eq.\,(\ref {Scons})
which like strangeness can in principle grow in hadronization,
 \begin{equation}  \label{encon}
 \left.{s\over S }\right|_{\mathrm H} \gtrsim
     \left.{s\over S }\right|_{\mathrm Q} .
\end{equation}
using   Eq.\,(\ref {Scons}) we underestimate the value of $ \gamma^{\mathrm{H}\,2}_q$.

Counting all
strange particles, the number of pairs is:
\begin{eqnarray}
\frac{dN_{s}^{\rm H}}{dy}=\frac{dV}{dy}\left[\right.
 &&\hspace*{-0.4cm}  {{\gamma^{\mathrm{H}}_{s}}}
  \left(\gamma^{\mathrm{H}}_qn^{\rm eq}_{K}
             +
      \gamma^{\mathrm{H}\,2}_qn^{\mathrm{eq}}_Y \right)\nonumber \\
+   && \hspace*{-0.4cm}
 \gamma^{\mathrm{H}\,2}_s(2\gamma^{\mathrm{H}}_qn^{\rm eq}_\Xi+n^{\rm eq}_\phi+P_sn^{\rm eq}_\eta) \nonumber \\
+  3 && \hspace*{-0.4cm}
        \gamma^{\mathrm{H}\,3}_sn^{\rm eq}_\Omega
\left. \right],
\label{gammas}
\end{eqnarray}
where $n^{\mathrm{eq}}_i$ are densities of strange hadrons
(mesons and baryons) calculated  using Eq.\,(\ref{dist}) in chemical
equilibrium. $P_s$ is the strangeness content of the $\eta$.
The way we count hadrons is to follow strangeness content, for  example
$n^{\rm eq}_K=n^{\rm eq}_{K^+}+n^{\rm eq}_{K^0}=n^{\rm eq}_{K^{-}}+n^{\rm eq}_{\bar{K}^0}$.
We impose  in our calculations $\bar{s}=s$.
The pattern of this  calculation follows an
established approach,  SHARE 2.1~\cite{Torrieri:2004zz} was  used in detailed evaluation.

\subsection{Strangeness per entropy $s/S$}
Considering that both strangeness, and entropy, are conserved in the
hadronization process, a convenient variable to consider as fixed
in the hadronization process, is the ratio
of these conserved quantities $s/S$. In chemical
equilibrium we expect that   in general such a ratio must be
different for different phases of matter from which particles
are produced~\cite{Kapusta:1986cb,Letessier:1993nz,Letessier:2006wn}.

\begin{figure}
\centering
\includegraphics[width=8cm,height=8cm]{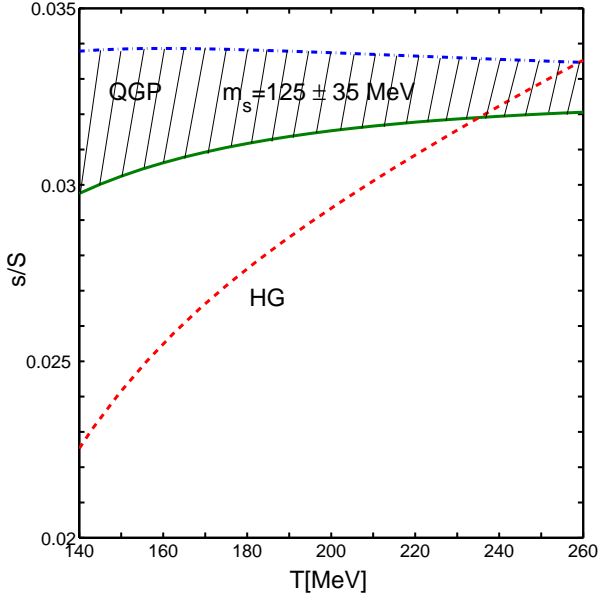}
\caption{(Color on line)
\small{Strangeness to entropy ratio $s/S$ as function of
temperature T, for the QGP (green, solid line for $m_s=160$ MeV,
blue dash-dot line for $m_s=90$ MeV) with $k=1$, see Eq.\,(\ref{str});
and for HG (light blue,dashed line) phases for
$\gamma_q=\gamma_s=\lambda_q=\lambda_s=1$
in both phases. }} \label{sS}
\end{figure}

We compare QGP and HG specific per entropy strange\-ness content
in figure {\ref{sS}}. We  show as function of temperature $T$
 the $s/S$ ratios for chemically equilibrated QGP and HG phase.
  For the QGP the entropy $S$ in QGP is calculated
as described in section \ref{entroSec}, and  we use $k=1$ in Eq.\,(\ref{str}).
 The shaded area
shows the range  of masses of strange
quarks, considered,  results for $m_s=90$ MeV (upper (blue) dash-dotted line)
and $m_s=160$ MeV ((green) solid line) form the boundaries. The central
QGP value is at about $s/S=0.032$.

The  short-dashed (light blue) line shows the hadron phase $s/S$ value found
using SHARE 2.1 . For HG  near to usual range of hadronization
temperature $T\simeq 160$\,MeV we find $s/S\simeq 0.025$. In general formation of
QGP implies and increase by 30\% in $s/S$. Both HG and QGP phases have a similar
specific strangeness content at $T=240$--260\,MeV, however it is
not believed that a HG at such high temperature would be a stable
form of matter.  This HG to QGP dissociation, or QGP hadronization depends on the
degree of strangeness equilibration in plasma~\cite{Rafelski:2005md},
and other dynamical factors.

In the QGP the value of $s/S$ for the range of realistic hadronization
temperature $140<T<180$ MeV is in general
 larger than in HG.  This implies that generally, the abundance of strange hadrons
produced in hadronization   over saturates   the strange
hadron  phase space, if QGP state had  reached (near) chemical  equilibrium. Moreover,
since we are considering the ratio $s/S$ and find in QGP a value greater than in
HG, for chemical equilibrium in QGP the hadronization process will lead to
$\gamma^{\rm H}_s/\gamma^{\rm H}_q>1$.

One can wonder if we have not overlooked some dynamical or microscopic
effect which could adjust the value of $s/S$ implied by QGP to the value
expected in HG. First we note
 that the fast growth of the volume $V$ cannot change $s/S$. Moreover, any
additional strangeness production in hadronization would enhance the
over-abundance  recorded in the resulting HG.  Only a highly significant entropy
production at fixed strangeness yield in the hadronization process could
bring the QGP $s/S$ ratio down, masking  strangeness over-saturation.
A  mechanism for such entropy production in hadronization
is unknown, and moreover, this  would further entail an
unexpected and high hadron multiplicity
excess.

One could of course argue that the perturbative QCD properties in the QGP are
meaningless, the entropy in QGP is much higher at given temperature.
However, the properties of QGP have been checked against the lattice
results, and the use of lowest order expressions is justified in
these terms~\cite{Letessier:2003uj}. Moreover, the value of $s/S$ is
established way before hadronization.

\subsection{Wr\'oblewski ratio $W_s$}
At this point it is appropriate to look  at another observable proposed
to study strangeness yield, the Wr\'oblewski ratio~\cite{Wroblewski:1985sz}:
\begin{equation}\label{WRs}
W_s\equiv \frac{2\langle \bar s s\rangle}{\langle \bar u u\rangle+\langle \bar d d\rangle}.
\end{equation}
$W_s$ compares the number of newly produced strange quarks to 
the produced number of light quarks. In an equilibrated   deconfined phase
$W_s$ compares the number of active strange quark 
degrees of freedom to the number of light quark degrees of freedom. 

The ratio $s/S$ compares the strange quark degrees of freedom to all degrees of 
freedom available in QGP. Therefore as function of $T$  the ratios  $s/S$ and $W_s$ can behave
differently: Considering the limit $T\to T_c$ a constant $s/S$ indicates 
that the reduction of $s$-degrees of freedom    goes hand in hand with  
the `freezing' of gluon degrees of freedom, which precedes the   `freezing' of light quarks.
This also  implies that for $T\to T_c$  in general $W_s$ diminishes.   
The magnitude of $m_s$, the strange quark mass decisively enters the  limit  $T\to T_c$.

For $T>> T_c$ the ratio $W_s$ 
can be evaluated comparing the rates of production of light and strange quarks,
using the fluctuation-dissipation theorem~\cite{Gavai:2002kq} , which allows 
to relate rate of quark production to 
 quark susceptibilities $\chi_i$  (see Eqs. (11) and (12) in~\cite{Gavai:2002kq}):
\begin{equation}\label{WRsc}
W_s\simeq R_\chi= \frac{2\chi_s}{\chi_u+\chi_d}.
\end{equation}
An evaluation of $R_\chi$ as function  of temperature  in lattice QCD  
has been  achieved~\cite{Gavai:2006ap}.  
For $T\simeq 2.5T_c$    the result obtained, 
 $  W_s\to R_\chi\simeq 0.8$,  is in agreement with the expectation for equilibrium QGP with 
nearly free quarks, with mass of strangeness having a small but noticeable  significance. 
With decreasing $T$, the ratio $R_\chi\to 0.3$ for  $  m_s=T_c$. However, this value
of $m_s$ is too large, the physical value should be nearly half as large, which  
would result in a greater $R_\chi$. Moreover, for $T\to T_c$ the 
relationship of $  W_s$ to $ R_\chi$, Eq. (\ref{WRsc})  is in question in that the greatly reduced 
rate of production of strangeness may not be
satisfying the conditions required in Ref.~\cite{Gavai:2002kq}.   

Comparing the observables $s/S$  and $W_s$ we note that the
experimental measurement requires in both cases a  detailed
analysis of   all particles produced. At lower reaction energies there
is additional complication in evaluation of $W_s$ due to the need to 
subtract the effect of quarks brought into the reaction region.  Turning
to the theoretical computation of   $s/S$  and $W_s$ we note 
that the thermal lattice QCD evaluation of $s/S$ is possible without any approximation, 
even if the actual computation of entropy near the phase boundary 
is a challenging task.  On the other hand, the lattice computation
of  $W_s$  relies on production rate of strangeness being sufficiently fast, 
which cannot be expected near to the phase boundary. Moreover, the 
variable    $s/S$ probes all QGP degrees of freedom, while  $W_s$ probes
only quark degrees of freedom. We thus conclude that  $s/S$ is both 
more accessible  theoretically and experimentally, 
and perhaps more QGP related  observable, as compared to $W_s$, 
since it comprises the gluon degrees of freedom.

\subsection{Strangeness chemical non-equilibrium}\label{noneqSec}
In order that in fast hadronization there is continuity of
strangeness   Eq.\,(\ref{sconshad}), and entropy, Eq.\,(\ref{encon})
the hadron phase  $\gamma^\mathrm{H}_s\ne 1$ and $\gamma^\mathrm{H}_q\ne 1$.
We have  to solve for  $\gamma^\mathrm{H}_s$ and  $\gamma^\mathrm{H}_q$
simultaneously  Eqs.\,(\ref{sconshad},\ref{encon}).

\begin{figure}  
\centering
\includegraphics[width=7.3cm,height=8.1cm]{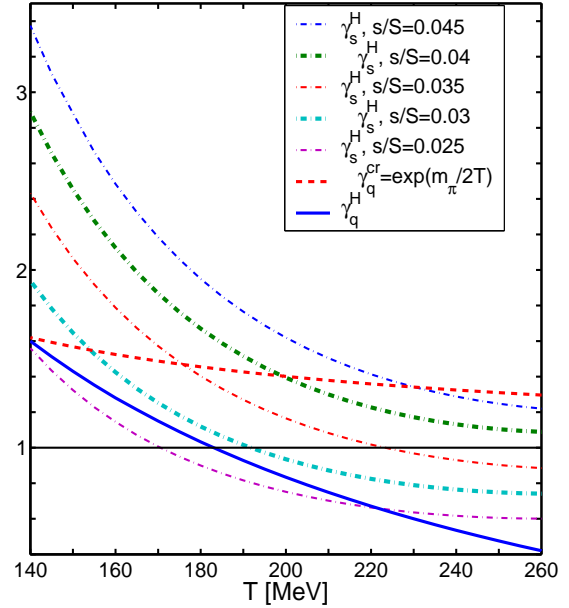}
\caption{(color on-line)
\small{Phase space occupancy as a function of $T$:
$\gamma^{\mathrm{H}}_q$ (blue, solid line),
$\gamma^\mathrm{H}_s$ (dash-dotted lines, from top to bottom)  for $s/S=0.045$,
for  $s/S=0.04$ (thick line), $s/S=0.035$, $s/S=0.03$ (thick line), $s/S=0.025$;
$\gamma^\mathrm{cr}_q$ (red, dashed line).}} \label{gq}
\end{figure}

In figure~{\ref{gq}} we show  as a function of $T$ the strange phase space occupancy
$\gamma^{\mathrm{H}}_s$,
obtained for several values of  $s/S$ ratio (from top to bottom 0.045, 0.04, 0.035, 0.03, 0.025)
evaluated for $S^{\rm Q}=S^{\rm H}$. The solid
line shows $\gamma^{\rm H}_q$  for $s^{\rm Q}=s^{\rm H}$ and $S^{\rm Q}=S^{\rm H}$.
The maximum allowed value Eq.\,(\ref{bcon}) is
shown dashed (red).

\begin{figure} 
\includegraphics[width=7.7 cm,height=14.2cm]{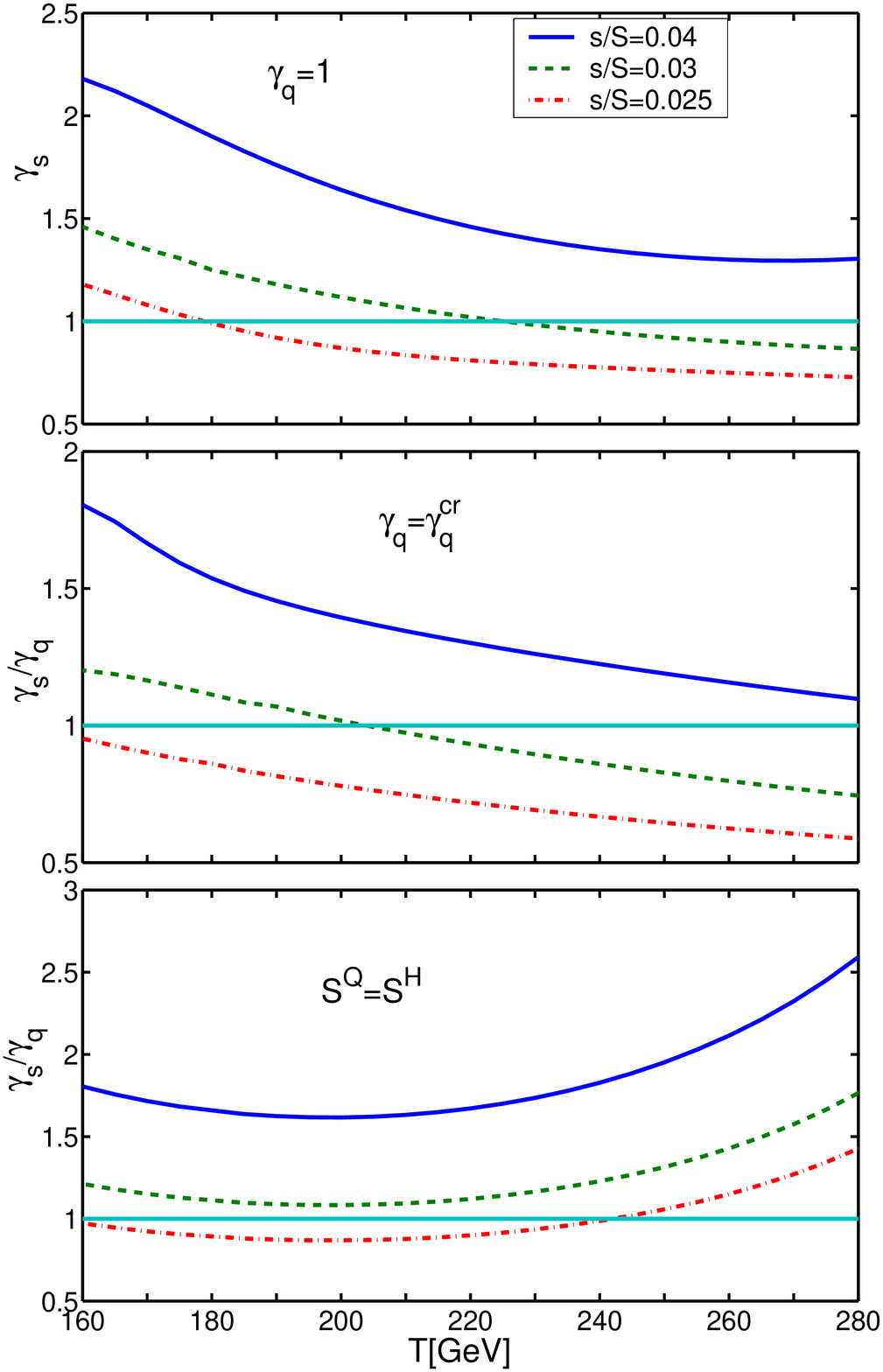}
\caption{(color on line)
\small{$\gamma^{\mathrm{H}}_s/\gamma^{\mathrm{H}}_q$ (=$\gamma^{\mathrm{H}}_s$  at
 $\gamma^{\mathrm{H}}_q=1$) as a function of
hadronization temperature $T$. Top frame:  $\gamma^{\mathrm{H}}_q=1$,
middle frame:   $\gamma^{\mathrm{H}}_q=\gamma^{cr}_q$, and bottom frame:
$S^H=S^Q$. Lines,  from top to bottom: $s/S=0.04$
(blue, solid line), $s/S=0.03$ (green, dashed line), $s/S=0.025$
(red, dash-dot line) }} \label{gseq} 
\end{figure}

In figure \ref{gseq} we show results for $\gamma^{\mathrm{H}}_s/\gamma^{\mathrm{H}}_q$
(where $\gamma^{\mathrm{H}}_q=1$ we show $\gamma^{\mathrm{H}}_s$ ) .
We consider the three cases: $\gamma_q=1$ , $\gamma^{\rm H}_q=\gamma^{cr}_q$,
and entropy conservation $S^H=S^Q$  for  $s/S=0.045$,
$s/S=0.04$, $s/S=0.035$, $s/S=0.03$, $s/S=0.025$ (dash-dot lines)
(lines from top to bottom).
We see that except  in case that strangeness were to remain well below
chemical equilibrium in QGP ($s/S\simeq 0.03$), the abundance of heavy
flavor hadrons we turn to momentarily will be marked by an
overabundance of strangeness, since practically
in all realistic conditions we find
$\gamma^{\rm H}_s>\gamma^{\rm H}_q$.

\begin{figure}
\centering
\includegraphics[width=9cm,height=9cm]{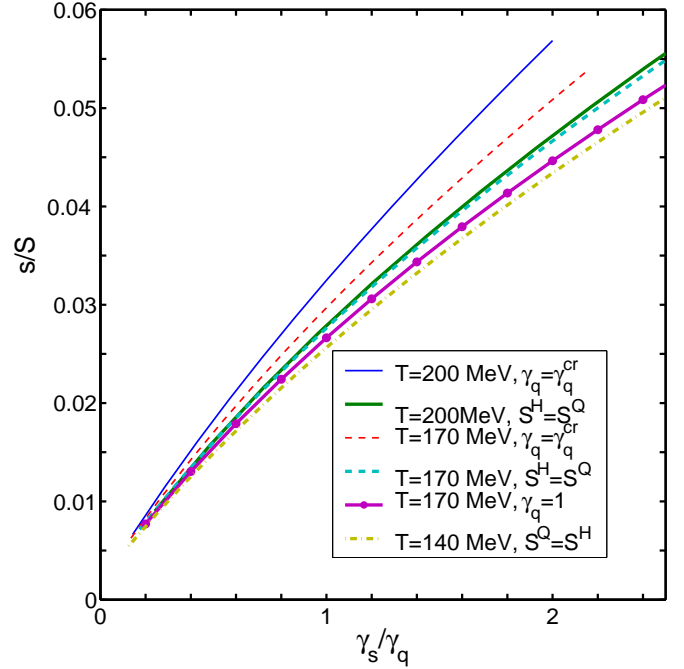}
\caption{(color on line) \small{Strangeness to entropy ratio, $s/S$,
as a function of $\gamma_s/\gamma_q$ for $T=200$ MeV,
$\gamma_q=0.083$, $S^H=S^Q$(solid line), $T=170$ MeV,
$\gamma_q=1.15$, $S^H=S^Q$ (dashed line), $T=140$ MeV,
$\gamma_q=1.6$, $S^H=S^Q$
(dash-dotted line); $\gamma_q = 1$ (dot
marked solid); $\gamma_q=\gamma_q^{cr}$: $T=200$ MeV (thin solid
line), $T=170$ MeV (thin dashed line).}}\label{sSrg}
\end{figure}

In figure~{\ref{sSrg}} we show  $s/S$ ratio as function of
$\gamma^{\rm H}_s/\gamma^{\rm H}_q$. The solid line is  for $T=200$
MeV, $\gamma^{\rm H}_q=0.83$, $S^Q=S^H$, dashed line for $T=170$
MeV, $\gamma^{\rm H}_q=1.15$  $S^Q=S^H$ and dash-dot line for
$T=140$, $\gamma^{\rm H}_q=1.6$ MeV, $S^Q=S^H$. We also consider
$\gamma_q=1$ case for $T=170$ MeV (dot marked (purple) solid line).
In this case strangeness content $\gamma_s/\gamma_q$ is higher than
for $S^H=S^Q$ with the same $T$ and $s/S$.  In the  limit
$\gamma^{\rm H}_q=\gamma_q^{cr}$, Eq.~\ref{bcon}, ($T=200$ MeV,
solid, thin line; $T=170$ MeV, dashed line) the strangeness content
$\gamma^{\rm H}_s/\gamma^{\rm H}_q$ is minimal for given $T$ and
$s/S$.

\begin{figure} 
\centering
\includegraphics[width=9cm,height=9cm]{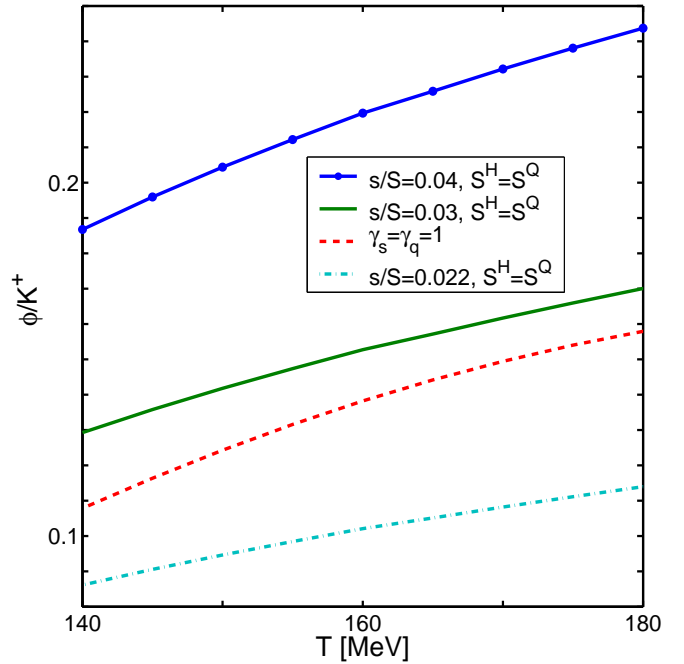}
\caption{(Color on line) The ratio  $\phi/{\rm K}^+$ as a function of $T$.
Dashed line (red) is for chemical equilibrium.
Solid line with dots (green)  $s/S=0.03$, solid line (blue)
$s/S=0.04$, dash-dot line (per)
is for $s/S=0.022$.} \label{phiK}
\end{figure}

These results suggest that it is possible to measure the value of $s/S$ irrespective
of what the hadronization temperature may be, as long as the main yield dependence is
on the ratio ${\gamma^{\rm{H}}_s}/{\gamma^{\rm{H}}_q}$.
Indeed,  we find that the ratio $\phi/K^+$;
\begin{equation}
 { \phi\over K^+} =\frac{\gamma^{\rm{H}}_s}{\gamma^{\rm{H}}_q}
                   \frac{ n^{\rm eq}_{\phi}}{n^{\rm eq}_{K^+}},
\label{gr}
\end{equation}
is less sensitive to hadronization temperature compared to its
 strong dependence on the value of $s/S$.
In figure~\ref{phiK} we show the total hadron phase space
 ratio $\phi/K^+$ as function of $T$  for several
$s/S$ ratios, and for $\gamma^{\rm{H}}_{s,q}=1$ (chemical equilibrium, dashed (red) line).
The $K^+$ yield contains the contribution from
the decay of $\phi$ into kaons which is a noticeable correction.

We record in table \ref{s/S} for given $s/S$ and volume $dV/dy$
the corresponding total yields of strangeness,
which may be a useful guide in consideration of the consistency of experimental
results with what we find exploring heavy flavor hadron abundance.

\begin{table}
\caption{Specific and absolute strangeness yield for different reaction volumes at $T=200$.
 \label{s/S}}
\begin{tabular}{|c|c|c|c|}
  \hline
    $s/S$&$ds/dy$&$dV/dy$ $[\mathrm fm^{-3}]$&T [MeV]\\
  \hline
$0.045$ &$550$ &$1000$& $200$\\
$0.04$ & $360$ &$800$& $200$\\
$0.035$ &$250$ &$700$& $200$\\
$0.03$ & $165$ &$600$& $200$\\
$0.025$& $106$ &$500$& $200$\\
$0.022$& $83$ &$500$& $200$\\
\hline
\end{tabular}
\end{table}

\section{Yields of heavy flavored hadrons}\label{heavyFlSec}
\subsection{Phase space occupancy $\gamma^{\mathrm{H}}_{c}$
and $\gamma^{\mathrm{H}}_{b}$}\label{gamcvalSec}
 The first step in order to determine the  yields of heavy flavor
hadronic particles is the determination of the phase space occupancy
$\gamma^{\mathrm{H}}_c$ and $\gamma^{\mathrm{H}}_b$.
$\gamma^{\mathrm{H}}_c$ is obtained from equality of
number of these quarks (i.e. of  quark and anti quark  pairs) in QGP
and HG.  The yield constraint is:
\begin{equation}
\frac{dN_{c}}{dy}=\frac{dV}{dy}\left[{\gamma^{\mathrm{H}}_{c}}n^{c}_{\mathrm{op}}
+
\gamma^{\mathrm{H}\,2}_{c}(n^{c\,eq}_{\mathrm{hid}}
+
2\gamma^{\mathrm{H}}_qn^{\mathrm{eq}}_{ccq}
+
2\gamma^{\mathrm{H}}_sn^{\mathrm{eq}}_{ccs})\right]; \label{gammacb}
\end{equation}
where open `op'  charm yield is:
\begin{equation}
n^{c}_{\mathrm{op}}\!=\gamma^{\mathrm{H}}_qn^{\mathrm eq}_{D}+\gamma^{\mathrm H}_sn^{\mathrm eq}_{Ds}+
{\gamma^{\mathrm{H}\,2}_q}\,n^{\rm eq}_{qqc}+{\gamma^{\mathrm{H}}_s}{\gamma^{H}_q}n^{\mathrm eq}_{sqc}+
{\gamma^{\mathrm{H}\,2}_s}\,n^{\mathrm{eq}}_{ssc}.
\end{equation}
Here $n^{\mathrm{eq}}_{D}$ and $n^{\mathrm{eq}}_{Ds}$ are densities of
$D$ and $D_s$ mesons, respectively, in chemical equilibrium, $n^{\mathrm{eq}}_{qqc}$
is equilibrium density of baryons with one charm and two light quarks,
$n^{\mathrm{eq}}_{ssc}$ is density of baryons with one
charm (or later on one bottom quark) and two strange quarks ($\Omega^0_c$,
$\Omega^0_b$) in chemical equilibrium and $n^{\rm eq}_{\mathrm{hid}}$
 is equilibrium particle density with both, a charm (or bottom) and an anticharm
(or antibottom) quark (C=0, B=0, S=0).
The equilibrium densities can be calculated using Eq.(\ref{dist}).
$\gamma^{\mathrm{H}}_c$ can now  be obtained from Eq.(\ref{gammacb}).

Similar calculations can be done for $\gamma^{\mathrm{H}}_b$. The
only difference is that we need to add number of $B_c$ mesons to the
right hand side of Eq.(\ref{gammacb}),
\begin{equation}
\frac{dN_{Bc}}{dy}=\gamma^{\mathrm{H}}_b\gamma^{\mathrm{H}}_c n^{\mathrm{eq}}_{Bc}
\frac{dV}{dy}.
\end{equation}
$n^{\mathrm{eq}}_{Bc}$ is density in chemical equilibrium of $B_c$.  In the
calculation of $\gamma^{\mathrm{H}}_c$ the contribution of term with $n_{Bc}$
is very small and we did not consider it above.

The value of $\gamma_c$ is in essence controlled by the open single
charm mesons and baryons. For this reason we do not consider the
effect of exact charm conservation. The relatively small effects due
to canonical phase space of charm are leading to a slight
up-renormalization of the value of $\gamma_c$ so that the primary
$dN_c/dy$ yield is preserved. This effect   enters into the yields
of multi-charmed and hidden charm hadrons, where the compensation is
not exact and there remains slight change in these yields. However,
the error made considering the high yield of charm is not important.
On the other hand for multi-bottom and hidden bottom hadrons the
canonic effect can be large, depending on actual bottom yield, and
thus we will not discuss in this paper yields of these hadrons,
pending extension of the methods here developed to include canonical
phase space effect.  



\begin{figure}
\centering
\includegraphics[width=9cm,height=12cm]{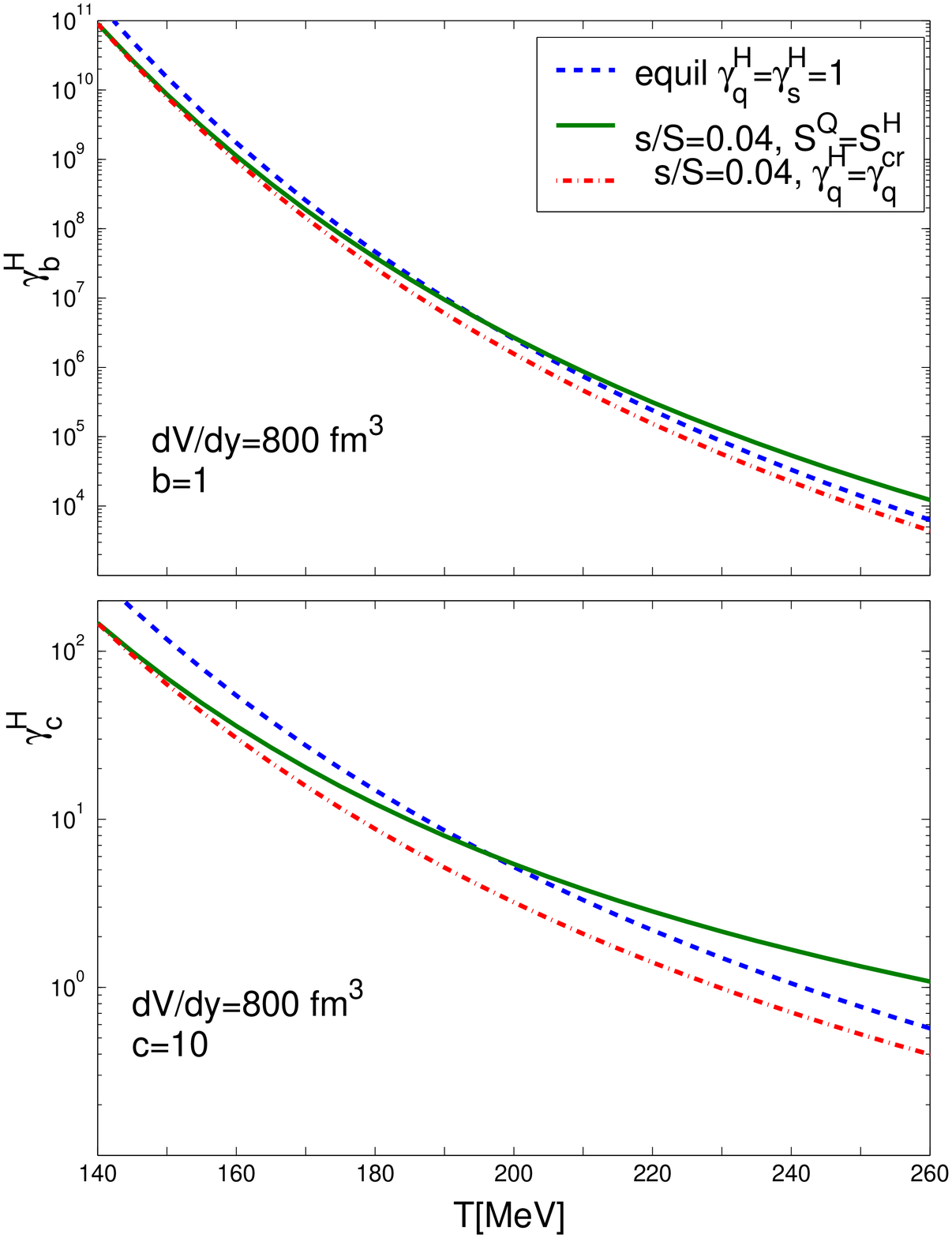}
\caption{(Color on line)
\small{$\gamma^{\mathrm{H}}_b (b=1)$ (upper panel),
and $\gamma^{\mathrm{H}}_c$ (c=10) (lower panel), as functions of
temperature of hadronization T. The solid lines are non-equilibrium for $s/S=0.04$ with $S^Q=S^H$,
dashed lines are
equilibrium case $\gamma_s$=$\gamma_q$=1 and dot-dash lines are for $s/S=0.04$ with maximal
value of $\gamma_q$($\gamma_q = \gamma^{cr}_q$) ($dV/dy=800$ $\mathrm{fm^3}$).}}
\label{gammaall}
\end{figure}

We consider in figure~\ref{gammaall} the temperature dependence of both
$\gamma^{\mathrm{H}}_b$ (top) and $\gamma^{\mathrm{H}}_c$ (bottom)
 for the heavy flavor yield given in Eqs.\, (\ref{nc},\ref{nb}).
In the non-equilibrium case (solid lines) the
space occupancy $\gamma^{\mathrm{H}}_s$ is obtained from
Eq.\,(\ref{gammas}) and $\gamma^{\mathrm{H}}_q$ is chosen to keep
$S^{\mathrm{H}}=S^{\mathrm{Q}}$.
$\gamma^{\mathrm{H}}_{c(b)}$
depend  on $\gamma_s$ and $\gamma_q$: the value of $N_s$ in Eq.(\ref{gammas})
is chosen to have $s/S=0.04$ after hadronization, the
corresponding $\gamma^{\rm H}_q$ and $\gamma^{\rm H}_s$ are shown
in figures~\ref{gq} and \ref{gseq} .
Since applicable  $\gamma^{\rm H}_q$ may depend on hadronization dynamics and/or details of  equation of state of
QGP, we show   charm quark phase space occupancies also for maximum possible
value of $\gamma^{\rm H}_q\to \gamma_q^{cr}$), also considered at $s/S=0.04$  for all hadronization temperatures.
We can compare our results with the chemical equilibrium (dashed lines)
setting $\gamma^{\mathrm{H}}_s = \gamma^{\mathrm{H}}_q=1$ in
Eq.\,(\ref{gammacb}). At hadronization condition $T=160\pm20$ MeV
temperatures we see in figure~\ref{gammaall} a significant (considering
the fast changing logarithmic scale) difference between the chemical equilibrium,
and non-equilibrium (s/S=0.04) results.

\begin{figure}
\centering
\includegraphics[width=8cm,height=8 cm]{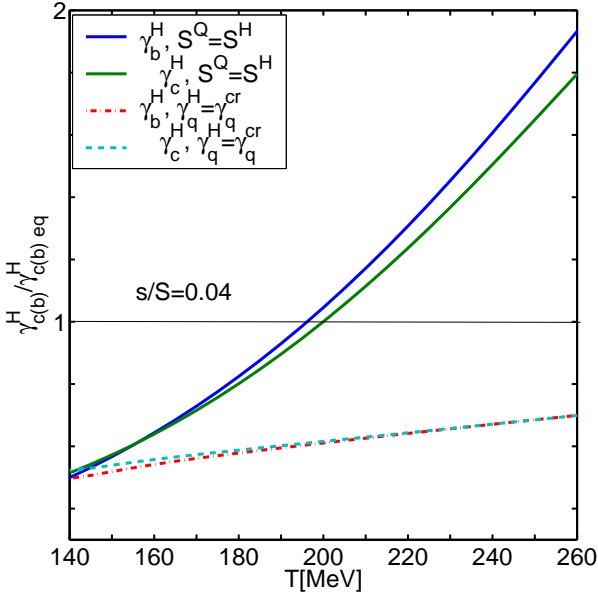}
\caption{(Color on line)
\small{$\gamma^{\mathrm{H}}_b/\gamma^{\mathrm{H}}_{b\,eq}$
and $\gamma^{\mathrm{H}}_c/\gamma^{\mathrm{H}}_{c\,eq}$, as functions of
temperature of heavy flavor hadronization T. The solid with dot marks line is
for $\gamma^{\mathrm{H}}_b/\gamma^{\mathrm{H}}_{b\,eq}$ with
$s/S=0.04$, solid line is for $\gamma^{\mathrm{H}}_c$ with $s/S=0.04$,
dot-dash and dashed lines are for $s/S=0.04$ with maximal value of
$\gamma_q\to \gamma^{cr}_q$  for $\gamma^{\mathrm{H}}_b/\gamma^{\mathrm{H}}_{b\,eq}$
and for $\gamma^{\mathrm{H}}_c/\gamma^{\mathrm{H}}_{c\,eq}$, respectively.}}
\label{rg}
\end{figure}

In figure~\ref{rg} we show  the ratio $\gamma^{\rm H}_{c(b)}/{\gamma^{\rm H}_{c(b)\,eq}}$ as a
function of hadronization temperature $T$. This helps us understand when the
presence of chemical nonequilibrium is most noticeable. This is especially
the case should heavy flavor hadronization occur at the same  temperature $T=140$--170 MeV
as is  obtained  for non-heavy  hadrons, and/or when the entropy content of light hadrons
is maximized with $\gamma_q^{\rm H}\to \gamma_q^{\rm cr}$.  When no additional entropy
is formed in hadronization, that is   $S^H=S^Q$,  $\gamma^{\rm H}_{c(b)}/{\gamma^{\rm H}_{c(b)\,eq}}$
exceeds unity for  $T>200$ MeV, at which point   the heavy flavor hadron yields
exceed the chemical equilibrium expectations.  In general we find that heavy hadron yields
if produced at normal hadronization temperature would be effectively suppressed, compared to
statistical equilibrium results, by the high strangeness yield. This happens since the phase space is
bigger at $\gamma^{\rm H}_{s,q}>1$, and thus a smaller $\gamma^{\rm H}_{c,b}$ is required to reach a given heavy flavor yield.

$\gamma^{\mathrm{H}}_b$ and $\gamma^{\mathrm{H}}_c$ are
nearly proportional to  $dN_{b,c}/dy$, respectively. The deviation
from the proportionality is due to the abundance of multi-heavy hadrons
and it is small. To estimate this effect more quantitatively we first evaluate:
\begin{equation}
\gamma^{\mathrm{H}}_{c0} =
\frac{dN_{c}}{dy}/\left(\frac{dV}{dy}n^c_{\mathrm{open}}\right),
\label{linsol}
\end{equation}
{\it i.e.\/} the value expected in absence of multi heavy hadrons.
Next we compare with the result when  we take into account
the last three terms in Eq.\,(\ref{gammacb}).
The influence of these therms depend not only
on $dN_c/dy$ but also on $dN_c/dy/dV/dy$.
For fixed $dV/dy=800$ $\mathrm{fm^{-3}}$ in the range of
$dN_c/dy=(5,30)$, we find that $\gamma^{\mathrm{H}}_c/N_c$
(and therefore yields of open charm hadrons)
changes at temperature $T=140$ MeV  by $\sim 6 \%$ for the $s/S=0.04$. For the chemical
equilibrium case $\gamma_s=\gamma_q=1$, $\gamma^{\mathrm{H}}_c/N_c$
changes up to $15 \%$ at the same conditions.
For the particles with hidden charm or 2 charm quarks the yields
are proportional $\gamma_i^2$,  therefore  changes in their yields will be
about twice larger. For RHIC $N_c<3$ and $dV/dy=600$ $\mathrm{fm^{-3}}$
the dependence of yields on $N_c$ is much smaller.

The multiplicity $dN_c/dy$ can also influence   $\gamma^{\mathrm{H}}_b$, since as
we noted it also includes a term
proportional to $\gamma^{\mathrm{H}}_cn^{\mathrm{eq}}_{Bc}$. In the
range of $N_c=(5,30)$, $\gamma_b/N_b$ changes at temperature
$T=0.14$ MeV by $\sim 0.5 \%$ for $s/S=0.04$. Since the mass of $b$-quark is much
larger than that of $c$-quark, the effect due to multi-bottom
states is negligible.

\subsection{D, Ds, B, Bs meson yields}\label{cbMesYielSec}
\begin{figure}
\centering
\includegraphics[width=7.3cm,height=9.8cm]{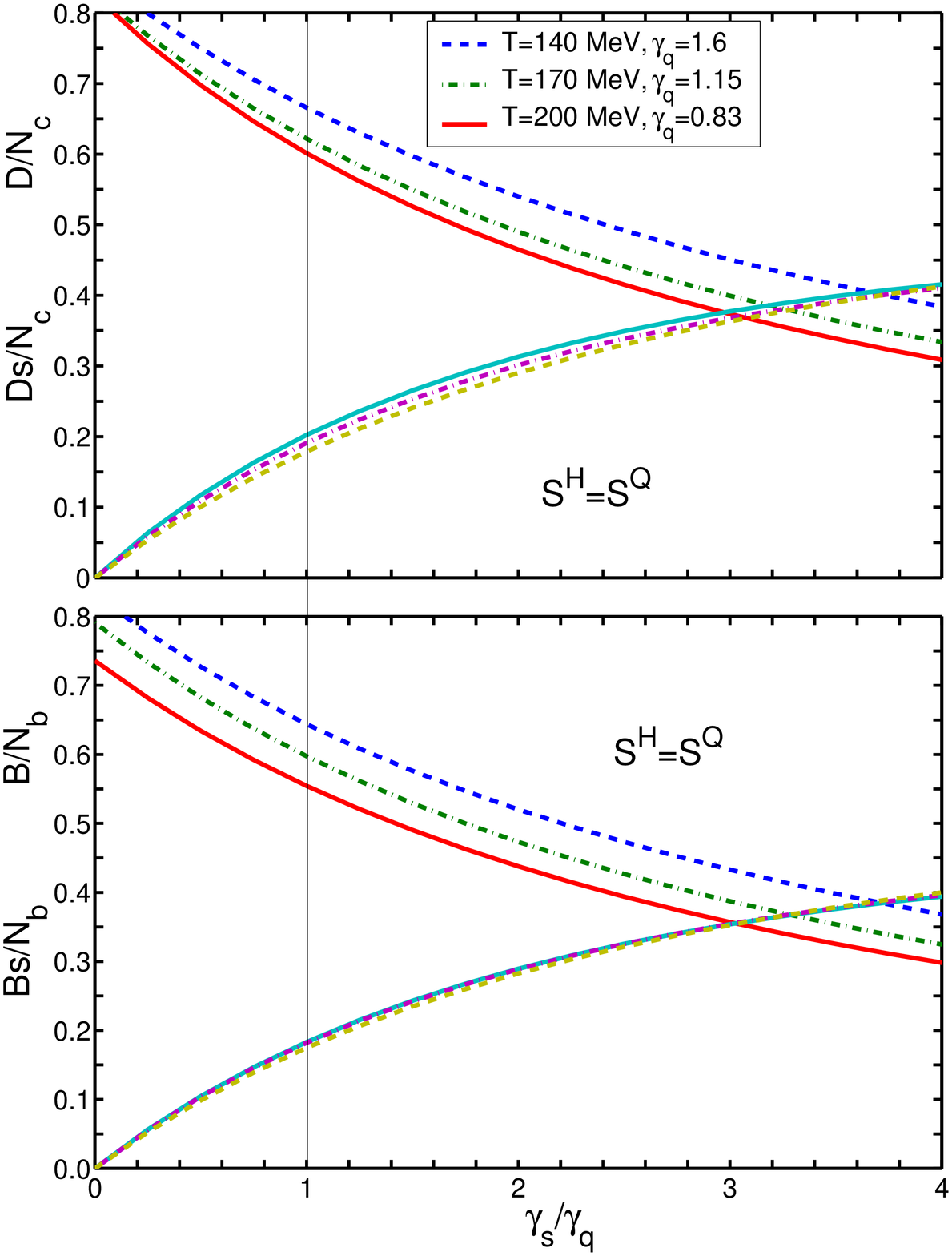}
\caption{(Color on line) \small{Upper panel, fractional charm meson
yield, and lower panel, fractional bottom meson yields as a function
of $\gamma_s/\gamma_q$ ratio for fixed hadronization temperature
$T$. Upper lines in each panel are for D(B) mesons, solid line is
for $T=200\,MeV$, $\gamma_q=1.1$, dashed line is for $T=170$ MeV,
$\gamma_q=1.15$ and dash-dot line is for $T=140$ MeV, $\gamma_q=0.83$ $(S^H=S^Q)$.
}} \label{mesrg}
\end{figure}

In next sections we will mostly consider particles yields after
hadronization and we will omit superscript H in $\gamma$s.
Considering Eq.\,(\ref{dist}),
 we   first obtain $\gamma_c$
as a function of $\gamma_s/\gamma_q$ ratio and T. Substituting this
$\gamma_c$ and appropriate  equilibrium hadron densities into
Eq.\,(\ref{dist}) we  obtain yields of $D(B)$ and $D_s(B_s)$, as
functions of the $\gamma_s/\gamma_q$ ratio, at fixed temperature,
which are shown on figure~\ref{mesrg}. In   the upper panel we show
the fractional yields of charmed $D/N_c$ and $D_s/N_c$ mesons, and
in the lower panel  $B/N_b$ and $B_s/N_b$ for $T=200$ MeV
(solid line), $T=170$ MeV (dashed line), $T=140$ MeV (dash-dot
line). Fractional yield means that these yields are normalized by
the total number of charm quarks $N_c$ and, respectively bottom
quarks $N_b$, and thus tell us how big a fraction of available heavy
flavor quarks binds to non-strange and strange heavy mesons,
respectively. Using figure~\ref{sSrg}  the  ratio
$\gamma_s/\gamma_q$ can be  related to the $s/S$ ratio. $\gamma_q$
was chosen to conserve entropy during hadronization process, see
figure~\ref{gq}. In general the heavy non-strange mesons yield
decreases and strange heavy meson yield increases with
$\gamma_s/\gamma_q$. The yields $D,B$ and $D_s,B_s$ are sum over
exited states of $D,B$ and $D_s,B_s$ respectively, see table
\ref{openbc} for the `vertical tower' of resonances we have
included.

Using $\gamma_c$, $\gamma_s$, $\gamma_q$ at a given $T$
(see figures~\ref{gq}, \ref{gseq}, \ref{gammaall}) we have now all the inputs
required to compute absolute and relative particle yields of all
heavy hadrons which we can consider  within the grand canonical
phase space. When we consider chemical equilibrium case, we  use
naturally $\gamma_s=\gamma_q=1$.

\begin{figure}
\centering
\includegraphics[width=9cm,height=12cm]{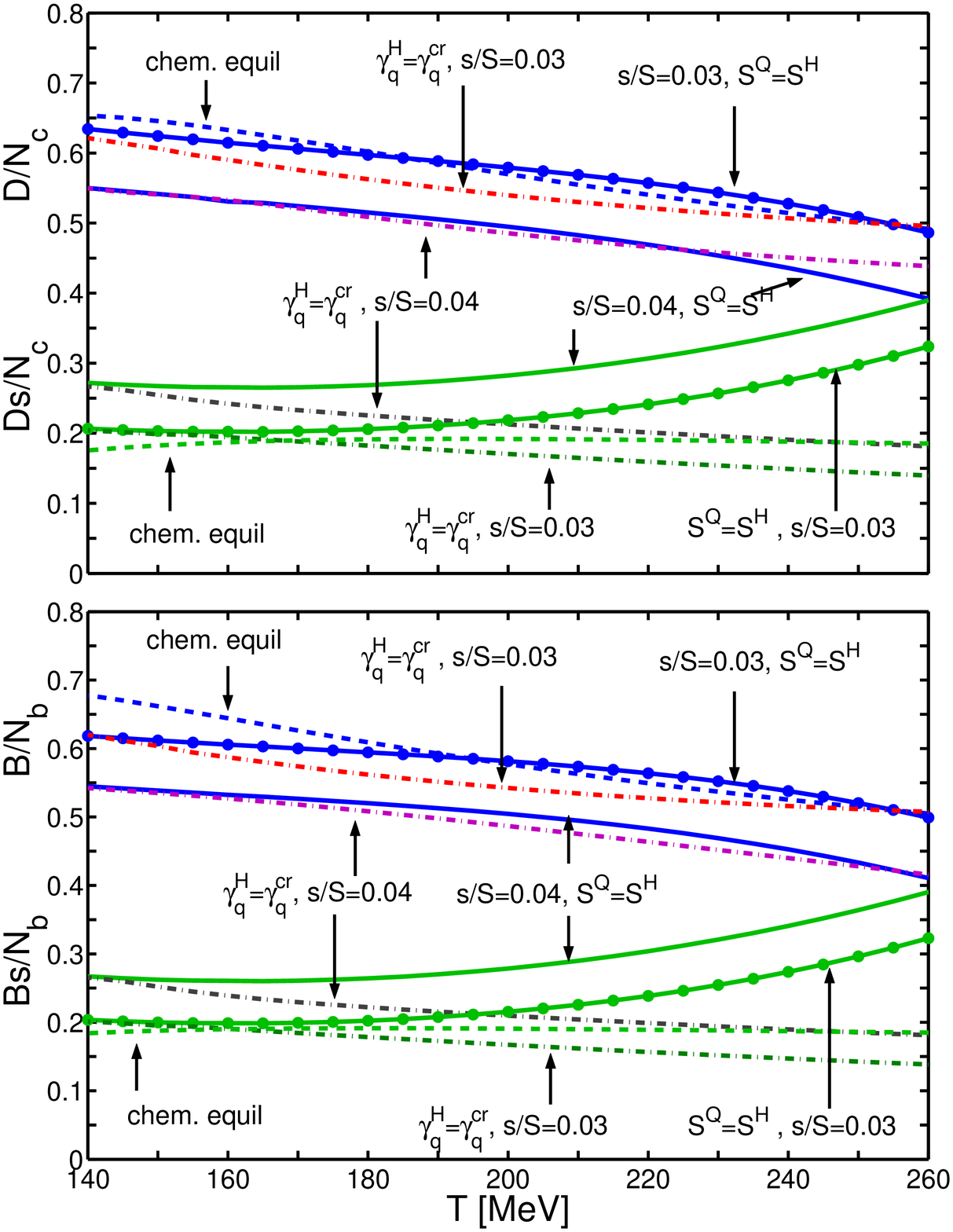}
\caption{(Color on line) \small{Upper panel, fractional charm meson
yield, and lower panel, fractional bottom meson yields. Equilibrium
(dashed lines) and non-equilibrium for $s/S=0.03$ (point marked
solid line) and $s/S=0.04$ (solid line) for $D/N_c$ (blue lines,
upper panel); $D_s/N_c$ (green lines, upper panel); for $D/N_c$ and
$D_s/N_c$ with $s/S=0.03$ and $s/S=0.04$ for $\gamma_q =
\gamma^{cr}_q$ (dash-dotted lines); $B/N_b$ (solid line, lower
panel); and $B_s/N_b$ (point marked solid line, lower panel), for
$B/N_b$ and $B_s/N_b$ with $s/S=0.03$ and $s/S=0.04$ for $\gamma_q =
\gamma^{cr}_q$ (dash-dot lines); as a function of $T$.}}
\label{Dmes}
\end{figure}

In figure~\ref{Dmes} we consider  the yields shown in
figure~\ref{mesrg} as a functions of hadronization temperature. The
dashed blue and green lines were obtained for chemical equilibrium
yields of $D$ and $D_s$ respectively. The extreme upper and lower
lines are for fractional $D$ and $D_s$ yields with $s/S=0.03$ (dot
marked, blue and green lines, respectively), while the central lines
are for $s/S=0.04$ (solid, blue and green lines). Also we show
fractional yields for maximal possible value $\gamma_q\to
\gamma^{cr}_q$, see figure~\ref{gq} for $\gamma^{cr}_q(T)$)
(dash-dot lines) and Eq.\,(\ref{bcon}).

\begin{figure}
\centering
\includegraphics[width=8.1cm,height=11.cm]{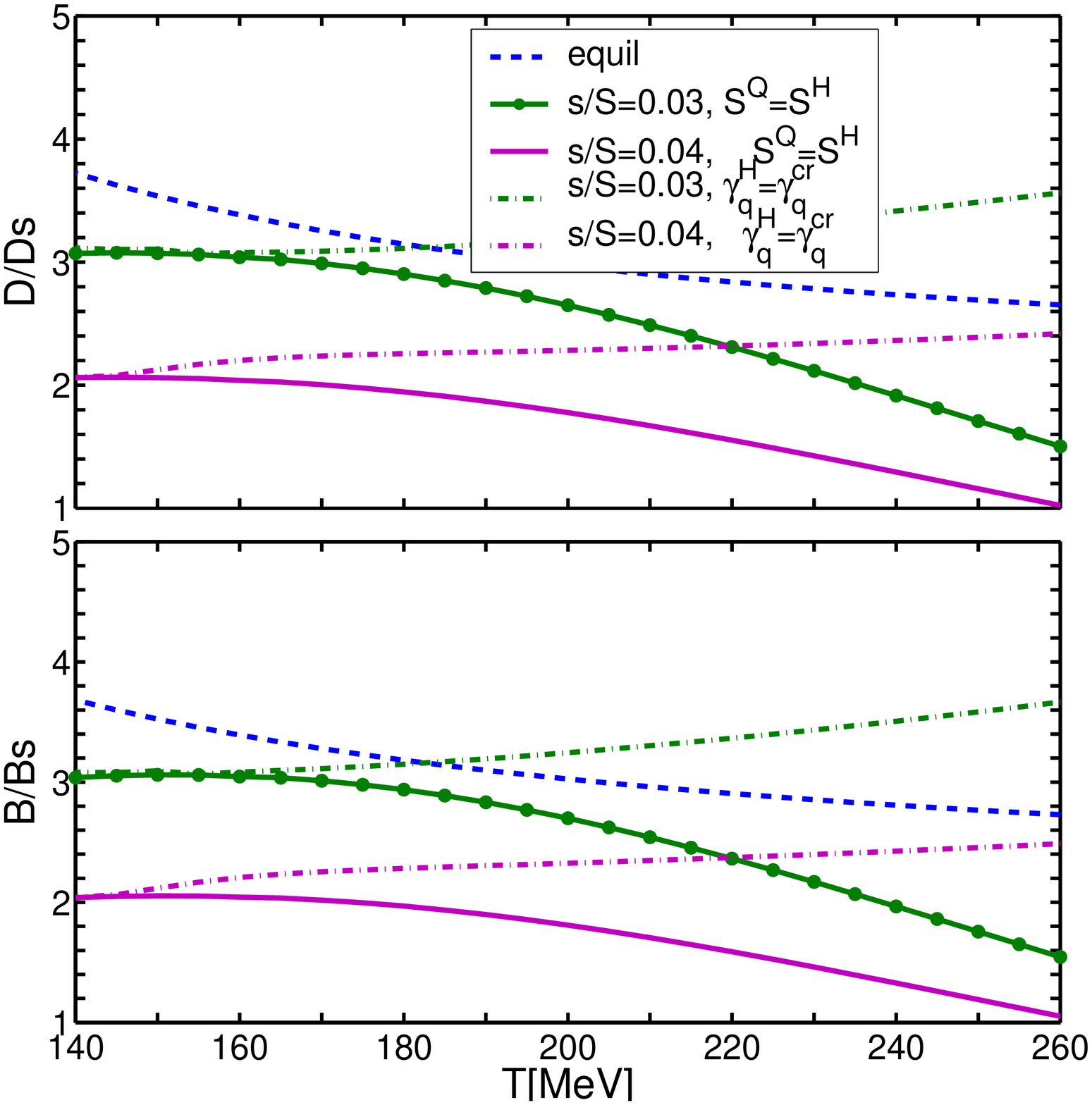}
\caption{(Color on line) \small{ Ratios $D/Ds$ (upper panel) and
$B/Bs$ (lower panel) are shown as a function of $T$ for different
$s/S$ ratios and in chemical equilibrium. Solid line is for
$s/S=0.04$, dash-dot line is for $s/S=0.03$, dashed line is for
$\gamma^{\mathrm{H}}_s=\gamma_s^{\mathrm{H}}=1$.}} \label{Dratio}
\end{figure}

We note that there is considerable symmetry at fixed $T$ between the
fractional yields of charmed, and bottom mesons,  for the same
condition of $s/S$. The chemical equilibrium results show
significant difference between strange and non-strange heavy mesons.
In the case of chemical equilibrium, for the considered very wide
range of hadronization temperatures  ${D_s}/N_c\simeq
{B_s}/N_b\simeq 0.2$ are nearly constant.
 A significant deviation from
this result would suggest the presence of chemical non-equilibrium
mechanisms of heavy flavor meson production.

The yields of $D_s/N_c$ and ${B_s}/N_b$ are very similar,
and similarly so for $D/N_c$ and $B/N_b$. Thus the relative
yield of either of these mesons measures the relative yield of charm to
bottom participating in the statistical hadronization process:
\begin{equation}
{D_s\over B_s}\simeq {D\over B}={N_c\over N_b}
\end{equation}
This is a very precise result, which somewhat depends on the
tower of resonances included, and thus in particular on the
symmetry in the heavy quark spectra between charmed and bottom states
which we imposed.

It is useful to reconsider here the ratio $D/D_s$ ($B/B_s$) which is
 proportional to $\gamma_q/\gamma_s$, see figure~\ref{rDsDg} for
$D_s/D$ presented as a function of $\gamma_s/\gamma_q$.
We consider this ratio now as a function of $T$, the upper panel
in figure \ref{Dratio} is for charm, the lower for bottom.
We see that there is considerable symmetry in the relative yields
between charmed and bottom mesons with upper and lower panels
looking quasi-identical. Except for accidental values of $T$ where
the equilibrium results (blue, dashed lines) cross the fixed $s/S$
results, there is considerable deviation in these ratios expected
from chemical equilibrium.
 For LHC with
$s/S=0.04$ this ratio is always noticeably smaller than in chemical equilibrium
(solid purple line is for $S^Q=S^H$ and purple, dash-dot line is for
$\gamma_q=\gamma_q^{cr}$). Even for
RHIC-like conditions with $s/S=0.03$   
this ratio is smaller than in chemical equilibrium for all temperatures when
entropy conservation in hadronization is assumed, $S^Q=S^H$ (dot marked
solid, green line).

\subsection{Heavy baryon yields}\label{BarYieSec}
\begin{figure}
\centering
\includegraphics[width=8.cm,height=12cm]{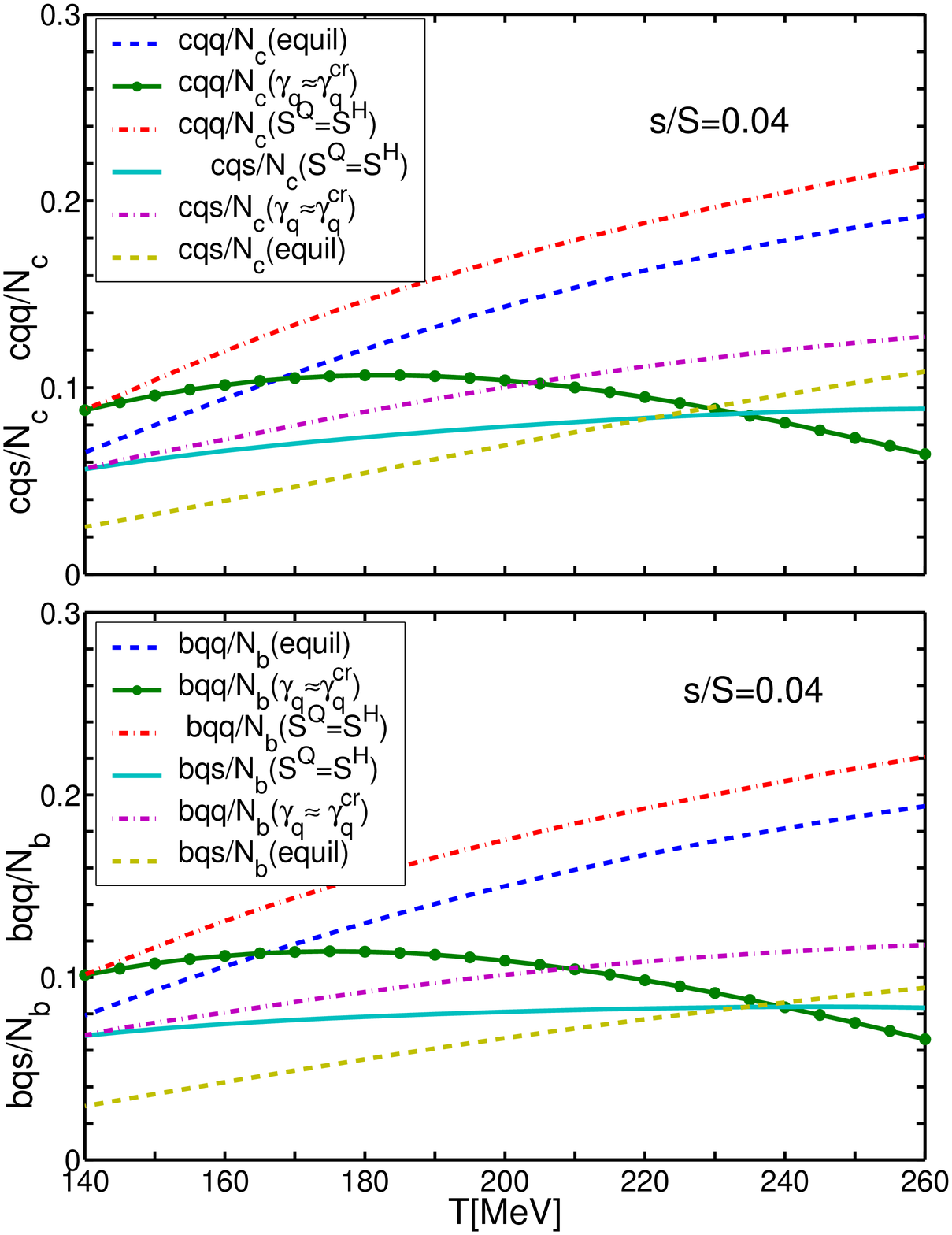}
\caption{(Color on line) \small{Equilibrium (dashed lines),
$s/S=0.04$, $S^Q=S^H$ (solid lines), $s/S=0.04$, $\gamma_q=\gamma^{cr}_q$, (the upper
panel) upper lines for each type are for ratio $(\Lambda_c+\Sigma_c)/N_c$ and lower lines are for $\Xi_c/N_c$ (upper panel) and
(lower panel) upper lines of each type are for $(\Lambda_b+\Sigma_b)/N_c$ and lower lines are for $\Xi_b/N_b$ as functions of T.}}
 \label{bar}
\end{figure}

As was the case comparing charm to bottom mesons we also establish a
symmetric set of charmed and bottom baryons, shown in the
table~\ref{baryonbc}. Many of the bottom baryons are result of theoretical
studies and we include that many states to be sure that both charm
and bottom are consider in perfect symmetry to each other. In
figure~\ref{bar} (upper panel) we show hadronization temperature
dependencies of yields of baryons with one charm quark normalized to
charm multiplicity $N_c$. We show separately yields of baryons
without strange quark $(\Lambda_c+\Sigma_c)/N_c$, and with one
strange quark S=1 ($\Xi_c/N_c$). We show two cases for
$s/S=0.04$ with conserved entropy at hadronization
$S^{\mathrm{Q}}=S^{\mathrm{H}}$ (solid lines) and with maximum
possible entropy value $\gamma_q=\gamma^{cr}_q$ (dash-dot lines).
The chemical equilibrium case $\gamma_q=\gamma_s=1$ is also shown
(dashed lines). The upper lines of each type are for
$(\Lambda_c+\Sigma_c)/N_c$, the lower lines are for $\Xi_c/N_c$.
A similar result is presented for bottom baryons in the lower panel of
figure~\ref{bar}. We note that the result for bottom baryons is more
uncertain since most baryon masses entering are not experimentally
verified.

\begin{table}
\caption{Charm and bottom baryon states  considered. States
in parenthesis are not known experimentally and have been
adopted from theoretical source~\cite{Albertus:2003sx}. \label{baryonbc}}
\begin{tabular}{|c|c|c|c|c|c|}
  \hline
  hadron&  \hspace*{-0.1cm}M[GeV] \hspace*{-0.2cm}&$Q:c,b$& hadron  & \hspace*{-0.1cm}M[GeV] \hspace*{-0.2cm}&g\\
\hline
  $\Lambda_c^+(1/2^+)$&2.285&udQ&$\Lambda_b0(1/2^+)$&5.624&2\\
  $\Lambda_c^+(1/2^-)$&2.593&udQ&$\Lambda_b0(1/2^-)$&(6.00)&2\\
  $\Lambda_c^+(3/2^-)$&2.627&udQ&$\Lambda_b0(1/2^-)$&(6.00)&2\\
  $\Sigma_c^+(1/2^+)$&2.452&qqQ&$\Sigma^0_b(1/2^+)$&(5.77)&6\\
  $\Sigma_c^{*}(3/2^+)$&2.519&qqQ&$\Sigma^{0*}_b(3/2^+)$&(5.78)&12\\
  $\Xi_c(1/2^+)$&2.470&qsQ&$\Xi_b(1/2^+)$&(5.76)&4\\
  $\Xi_c^{'}(1/2^+)$&2.574&qsQ&$\Xi^{'}_b(1/2^+)$&(5.90)&4\\
  $\Xi_c(3/2^+)$&2.645&qsQ&$\Xi^{'}_b(3/2^+)$&(5.90)&8\\
  $\Omega_c(1/2^+)$&2.700&ssQ&$\Omega_b(1/2^+)$&(6.00)&2\\
  $\Omega_c(3/2^+)$&(2.70)&ssQ&$\Omega_b(3/2^+)$&(6.00)&4\\
  \hline
\end{tabular}
\end{table}

We note that the results shown  figure~\ref{bar} imply
that under LHC conditions at least  15\% of heavy flavor can
be bound in heavy baryons, but possibly 30\%. For large
$\gamma_q=\gamma^{cr}_q > 1$ we see increase in
$(\Lambda_c+\Sigma_c)/N_c$ yields compared to chemical equilibrium
and especially compared to  entropy conserved hadronization $S^Q=S^H$.
This is so since    yields are
proportional to $\gamma_q^2$, $\gamma_s \gamma_q $. This results to
relative suppression the $D_s/N_c$ (see figure~\ref{Dmes}).

\begin{figure}
\centering
\includegraphics[width=7.2cm,height=7.2cm]{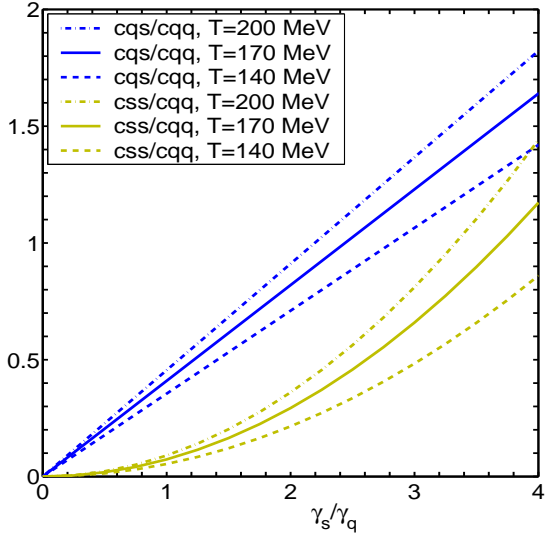}
\caption{(Color on line)
\small{The ratios $cqs/cqq=\Xi_c/(\Lambda_c+\Sigma_c)$ (upper lines)
and $css/cqq=\Omega_c/(\Lambda_c+\Sigma_c)$ (lower lines)
for $T=200 MeV$ (dash-dot line), $T=170$ MeV (solid line) and
$T=140$ MeV (dashed line) as functions of $\gamma_s/\gamma_q$} .}
\label{barratio}
\end{figure}

In figure~\ref{barratio} we show ratio
$cqq/cqs=(\Lambda_c+\Sigma_c)/\Xi_c$ as a function of
$\gamma_s/\gamma_q$ for $T=200$ MeV (dash-dot line),
$T=170$ MeV (solid line) and $T=140$ MeV (dashed line).
This dependence is linear,  the slope   depends only on
hadronization temperature $T$.  The $\gamma_s/\gamma_q$ ratio can be converted to $s/S$ ratio using figure~\ref{sSrg}.

\begin{figure}
\centering
\includegraphics[width=8cm,height=8cm]{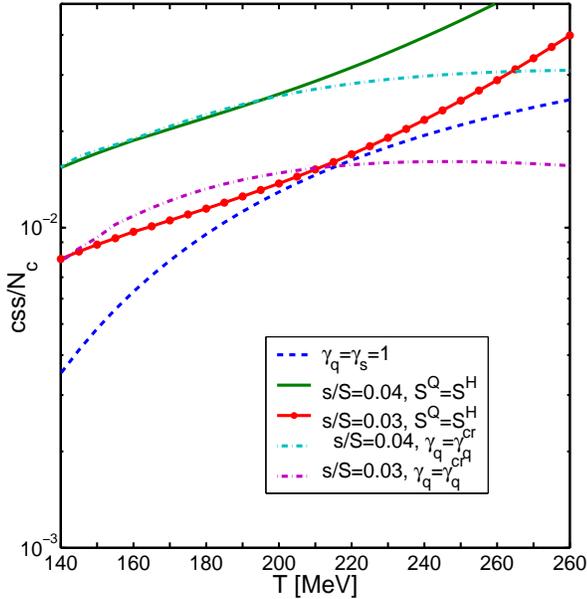}
\caption{(Color on line)
\small{$\Omega_c(css)/N_c$ as function of $T$: dashed line for chemical equilibrium;
solid lines are
 for $S^{\mathrm{Q}}=S^{\mathrm{H}}$, dashed dotted lines are for $\gamma_q=\gamma^{cr}_q$: both for
$s/S=0.03$  and $s/S=0.04$ (upper lines).}}
\label{barss}
\end{figure}

The yield  of multi-strange charmed baryon, $\Omega_c(css)$
is, similar to the light multi-strange hadrons, much more sensitive
to chemical non-equilibrium. In figure~\ref{barss} we see a
large increase in fractional yield of $\Omega_c(css)/N_c$
for $s/S=0.04$ and $S^{\mathrm{Q}}=S^{\mathrm{H}}$ (solid line) compared
to the chemical equilibrium (dashed line) expectation
for the entire considered range of hadronization  temperature. As expected,
this yields increase  with $T$. This also means that higher formation
temperature can be invoked to explain an unusually high yield.
We expect that at LHC more than one percent of total
charm yield will  be found in the $\Omega_c(css)$  state.

\subsection{Yields of hadrons with two heavy quarks}\label{MultiSec}
We consider multi-heavy hadrons listed in the table~\ref{multihiddennbc}.
The yields we will compute are now more model dependent since
we cannot completely reduce the result, it either remains dependent on
the reaction volume $dV/dy$, or on the total charm(bottom) yields $dN/dy$.
For example the yields of hadrons  with two heavy quarks are
approximately proportional to $1/(dV/dy)$ because $\gamma^H_{b,c}$ for heavy
quarks is proportional to $1/dV/dy$, see Eq.\,(\ref{gammacb}):
\begin{eqnarray}
\frac{dN_{hid}}{dy} &\propto& \gamma^{H\,2}_c \frac{dV}{dy} \propto
\frac{1}{dV/dy},\\
\frac{dN_{Bc}}{dy} &\propto& \gamma^H_c \gamma^H_b \frac{dV}{dy}
\propto \frac{1}{dV/dy}.
\end{eqnarray}
Moreover, unlike it is the case for single heavy hadrons,  the
canonical correction to grand-canonical phase space does not cancel out
in these states, adding to the uncertainty.

\begin{table}
\caption{Hidden charm and multi heavy hadron states
considered. States in parenthesis are not known experimentally
\label{multihiddennbc}}
\begin{tabular}{|c|c|c|c|}
  \hline
  hadron&  &mass(GeV)&g\\
  \hline
   $\eta_c (1S)$&$c\bar{c}$&2.9779&1\\
   $J\!/\!\Psi (1S)$&$c\bar{c}$&3.0970&3\\
   $\chi_{c0}(1P)$&$c\bar{c}$ &3.4152&1\\
   $\chi_{c1}(1P)$&$c\bar{c}$ &3.5106&3\\
   $h_c(1P)$&$c\bar{c}$ &3.526&3\\
   $\chi_{c2}(1P)$&$c\bar{c}$ &3.5563&5\\
   $\eta_c(2S)$&$c\bar{c}$&3.638&1\\
   $\psi(2S)$&$c\bar{c}$&3.686&3\\
   $\psi$&$c\bar{c}$&3.770&3\\
   $\chi_{c2}(2P)$&$c\bar{c}$&3.929&5\\
   $\psi$&$c\bar{c}$&4.040&3\\
   $\psi$&$c\bar{c}$&4.159&3\\
   $\psi$&$c\bar{c}$&4.415&3\\
   $B_c$&$b\bar{c}$&6.27&1\\
   $\Xi_{cc}$&ccq&3.527&4\\
   $\Omega_{cc}$&ccs&(3.660)&2\\
\hline
\end{tabular}
\end{table}

\begin{figure}
\centering
\includegraphics[width=8cm,height=10cm]{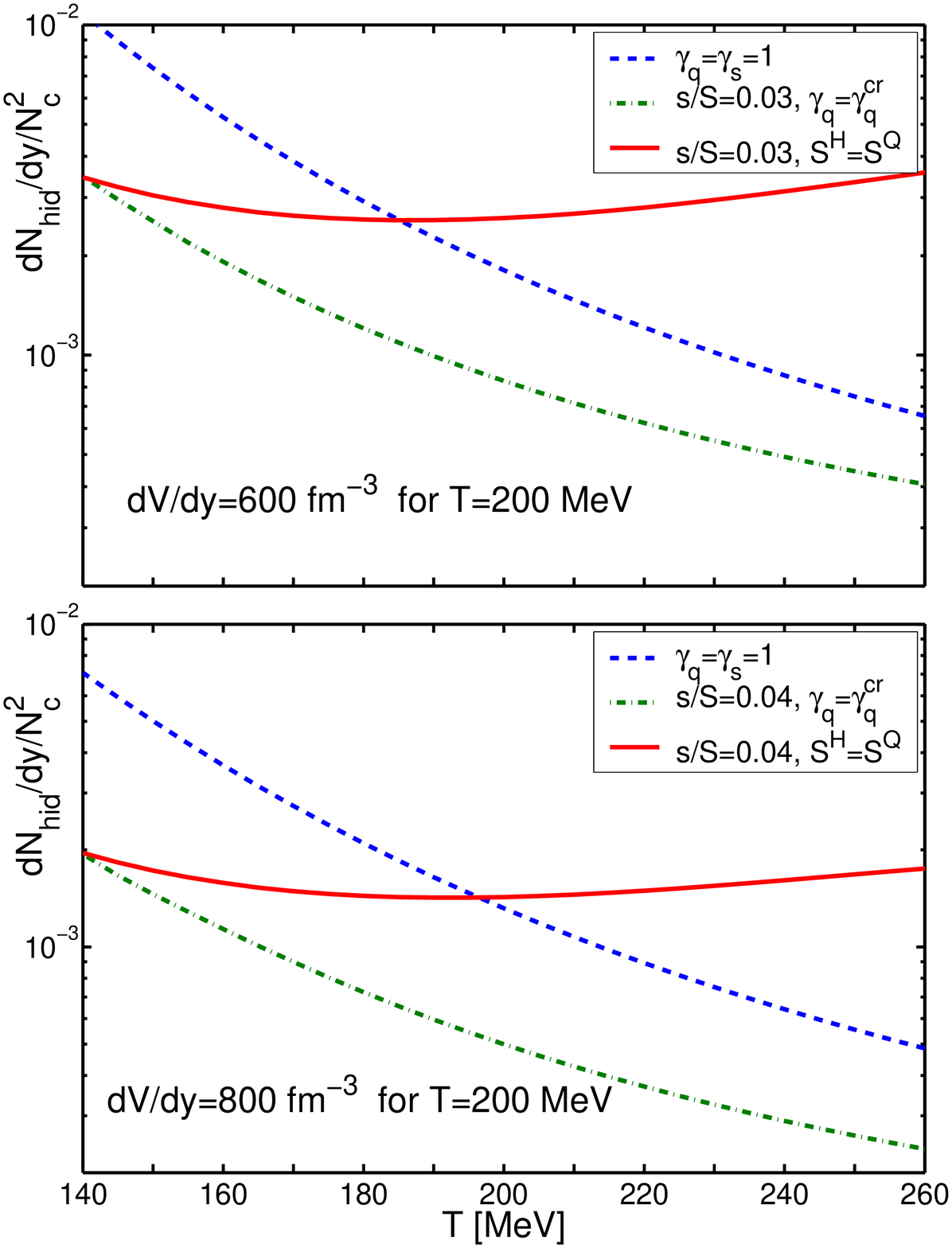}
\caption{(Color on line) \small{$c\bar{c}/N^2_c$  yields as a
function of hadronization temperature T, at
  $dV/dy=600$ $\mathrm{fm^{-3}}$ for $T=200$ MeV, $s/S=0.03$  (upper panel),
$dV/dy=800$ $\mathrm{fm^{-3}}$ for $T=200$ MeV,  $s/S=0.04$ (lower panel).
Results shown are for  $S^Q=S^H$  (solid lines),   for $\gamma_q=\gamma_q^{cr}$  (dash-dot lines),
and    for chemical equilibrium case (dashed lines, $s/S$ is not fixed).}} \label{cc}
\end{figure}

Thus the result we present must seen as a guiding the eye and
demonstrating a principle.
In figure~\ref{cc} we show the yield of hidden charm $c\bar{c}$ mesons
(see table~\ref{multihiddennbc}) normalized by the  square
of charm multiplicity $N_c^2$ as a
function of hadronization temperature $T$. We consider again cases with
$s/S=0.03$ (upper panel) and $s/S=0.04$ (lower panel),
 solid line is for $S^H=S^Q$, dot-dash line is for $\gamma_q=\gamma^{cr}_q$,
and dot-dash line is for $\gamma_q=\gamma_q^{cr}$.
The chemical equilibrium $c\bar{c}$ mesons yields are
shown (dashed lines on both panels) for two different values of
$dV/dy=600$\,fm$^3$ for $T = 200$ MeV (upper panel) and
$dV/dy=800$\,fm$^3$ for $T = 200$ MeV (lower panel).

The yield of $c\bar{c}$ mesons is much smaller for $s/S=0.04$ than
in equilibrium for the same $dV/dy$ for large range of hadronization
temperatures. For $s/S=0.03$ the effect is similar, but suppression
is not as pronounced. For $\gamma_q=\gamma_q^{cr}$ suppression
the yield of hidden charm particles is always
smaller than equilibrium value. This suppression occurs due to
competition with the yield of strange-heavy mesons,
and also, when $\gamma_q>1$, with
heavy baryons with two light quarks.
The enhanced yield of $D, D_s$ and heavy baryons in effect depletes
the pool of available charmed quark pairs, and fewer hidden charm
$c\bar{c}$ mesons are formed. For particles with two heavy quarks
the effect is larger than for hadrons with one heavy quark and light
quark(s).

\begin{figure}
\centering
\includegraphics[width=9cm,height=9cm]{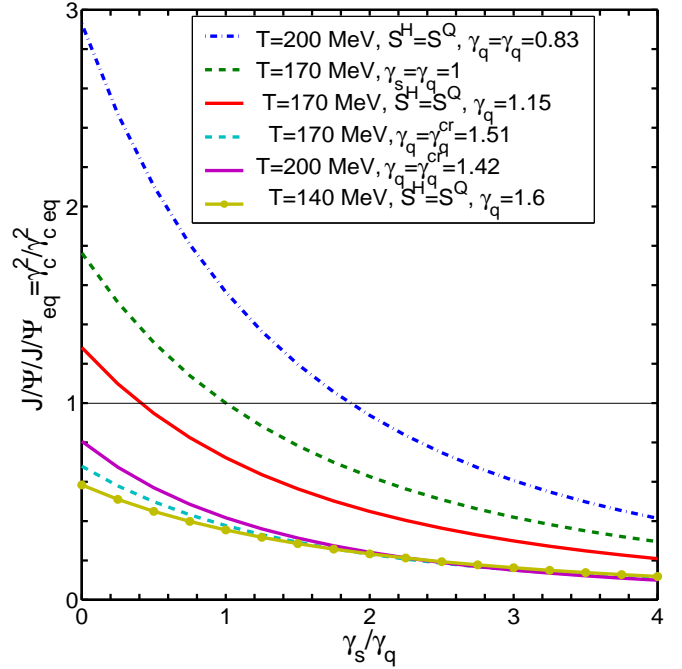}
\caption{(Color on line) \small{Ratio $J\!/\!{\Psi}/J\!/\!\Psi_{eq}=\gamma^2_c/\gamma^2_{c\,eq}$ as a function
of $\gamma^{\rm H}_s/\gamma^{\rm H}_q$ at fixed value of $\gamma^{\rm H}_q$ and if required, entropy
conservation.
Shown are:   $T=200$ MeV at $\gamma_q=0.83$ (dot-dash line) and at
$\gamma_q=\gamma^{cr}_q=1.42$ (lower solid line (purple) );
$T=170$ MeV at $\gamma_q=1$(upper dashed line) , at  $\gamma_q=1.15$, (upper solid line (red)), and at
$\gamma_q=\gamma^{cr}_q=1.51$, (lower dashed line); and
 $T=140$ MeV, $\gamma_q=1.6$}} \label{jpsrg}
\end{figure}

In  figure~\ref{jpsrg} we compare the $J\!/\!\Psi$ yield to the chemical equilibrium
yield $\Psi/J\!/\!\Psi_{eq}$, as a function of $\gamma^{\rm H}_s/\gamma^{\rm H}_q$,
each line is at a fixed value $\gamma^{\rm H}_q$.
This ratio is:
\begin{equation}
\frac{J\!/\!{\Psi}}{J\!/\!{\Psi}_{\rm eq}}=\frac{N_{hid}}{N_{hid\,eq}}=\frac{\gamma_c^2}{\gamma_{c\,{\rm eq}}^2}.
\end{equation}
$J\!/\!{\Psi}/J\!/\!{\Psi}_{\rm eq}$  always decreases when $\gamma_s/\gamma_q$ increases.
For $\gamma_q = \gamma_{cr}$ $J\!/\!\Psi/J\!/\!\Psi_{\rm eq}$
is smaller than unity  even when $\gamma_s\to 0$, because of large phase space occupancy of light quarks.
$J\!/\!{\Psi}/J\!/\!{\Psi}_{\rm eq}>1 $ for small $\gamma_q$ and small $\gamma_s/\gamma_q$ .

\begin{figure}
\centering
\includegraphics[width=8cm,height=8cm]{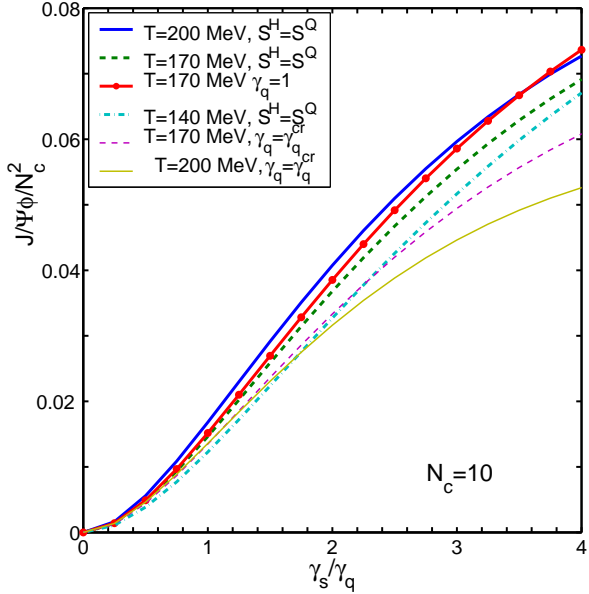}
\caption{(Color on line) \small{$J/\Psi\phi/N^2_c$ states yields as a
function of $\gamma_s/\gamma_q$ ratio for $T=200$ MeV, $S^Q=S^H$ (solid line)
and $\gamma_q=\gamma_q^{cr}$ (solid thin line),
$T=170$ MeV: $S^Q=S^H$ (dashed line), $\gamma_q=\gamma_q^{cr}$ (thin dashed line)
 and $\gamma_q=1$ (solid line with dot marker); for
$T=140$ MeV, $S^Q=S^H$ (dash-dot line)}} \label{jpsphrg}
\end{figure}

 Considering  the product of $J\!/\!\Psi$ and $\phi$
yields normalized by $N^2_c$ we eliminate nearly all the uncertainty
about the yield of charm and/or hadronization volume. However,
we tacitly assume that both $J\!/\!\Psi$ and $\phi$ hadronize at the
same temperature.  In figure~\ref{jpsphrg} we show $J/\Psi\phi/N^2_c$
as function of $\gamma_s/\gamma_q$. There is considerable  difference
to the ratio considered in figure~\ref{JpsiD}.  We see mainly dependence on
 $\gamma_s/\gamma_q$. As before, see
 section \ref{singHadSec} $J\!/\!\Psi$ is the sum of all states
$c\bar{c}$ from table~\ref{multihiddennbc} that can decay to
$J\!/\!\Psi$. We show results for $T=200$
MeV (solid lines), $T=170$ MeV (dashed line) and $T=140$ MeV
(dash-dot line). The $\gamma_q$, for each $T$,  is fixed by   entropy
conservation condition during hadronization (figure~\ref{gq}) (thick lines) or by
$\gamma_q=\gamma_q^{cr}$ (thin lines). For $T=140$ MeV these lines
coincide.   $T=170$ MeV, $\gamma_q=1$ case is also shown (solid
line with dot markers). The  $s/S$ values, which correspond  to given
$\gamma_s/\gamma_q$ ratio can be found in figure~\ref{sSrg}.
Figure~\ref{jpsphrg} shows  that despite the yield $\phi/(dV/dy)$ increasing as $(\gamma_s/\gamma_q)^2$,
$J\!/\!{\Psi}\phi/N^2_c$ is increasing as $\gamma_s/\gamma_q$ considering
compensation effects.

\begin{figure}
\centering
\includegraphics[width=8cm,height=12cm]{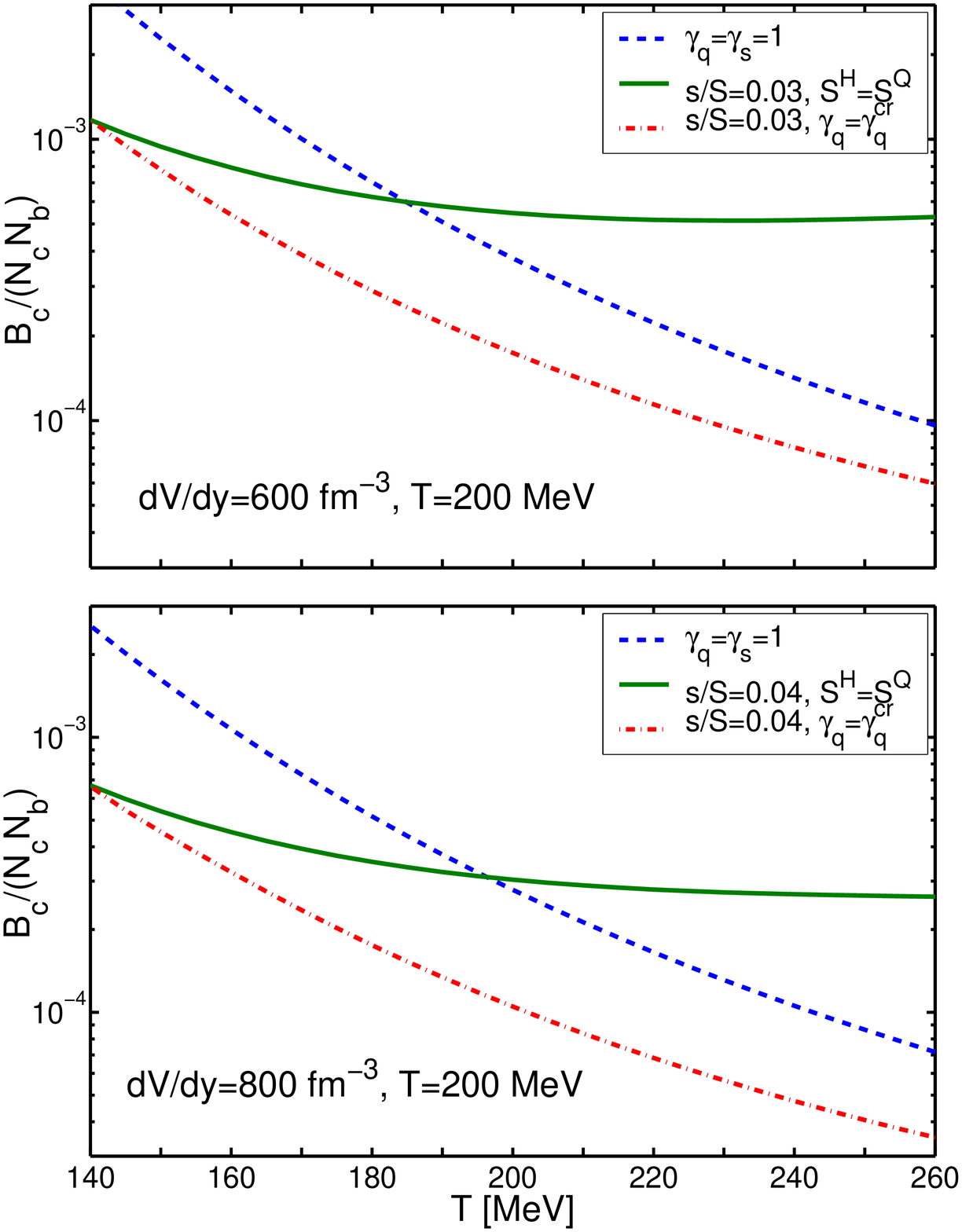}
\caption{(Color on line) \small{Bc mesons yields as function of T for
chemical equilibrium case with $dV/dy=600$ $fm^{-3}$ for $T=200$ MeV
(the upper panel, dashed line), for $s/S=0.03$ with $dV/dy=600$
$fm^{-3}$ for $T=200$ MeV (the upper panel, solid line), for
chemical equilibrium case with $dV/dy=800$ $fm^{-3}$ for $T=200$
MeV (the lower panel, dashed line) for $s/S=0.04$ $dV/dy=800$
$fm^{-3}$ for $T=200$ MeV (lower panel, solid line)}} \label{bc}
\end{figure}

A similar situation,
as in figure~\ref{jpsrg} for hidden charm,
arises for the $B_c$ meson yield, see
figure~\ref{bc}, where $B_c/N_cN_b$ ratio is shown as a function of
hadronization temperature $T$, for the same strangeness yield cases
as discussed for the hidden charm meson yield.  Despite
suppression in strangeness rich environment, the  $B_c$ meson yield
continues to be larger than the yield of $B_c$ produced in single NN
collisions, where the scale yield is at the level of
$\sim{10^{-5}}$, see cross sections for $b\bar{b}$ and $B_c$
production in~\cite{Anikeev:2001rk} and in \cite{Chang:2003cr}, respectively.

\begin{figure}
\centering
\includegraphics[width=8cm,height=12cm]{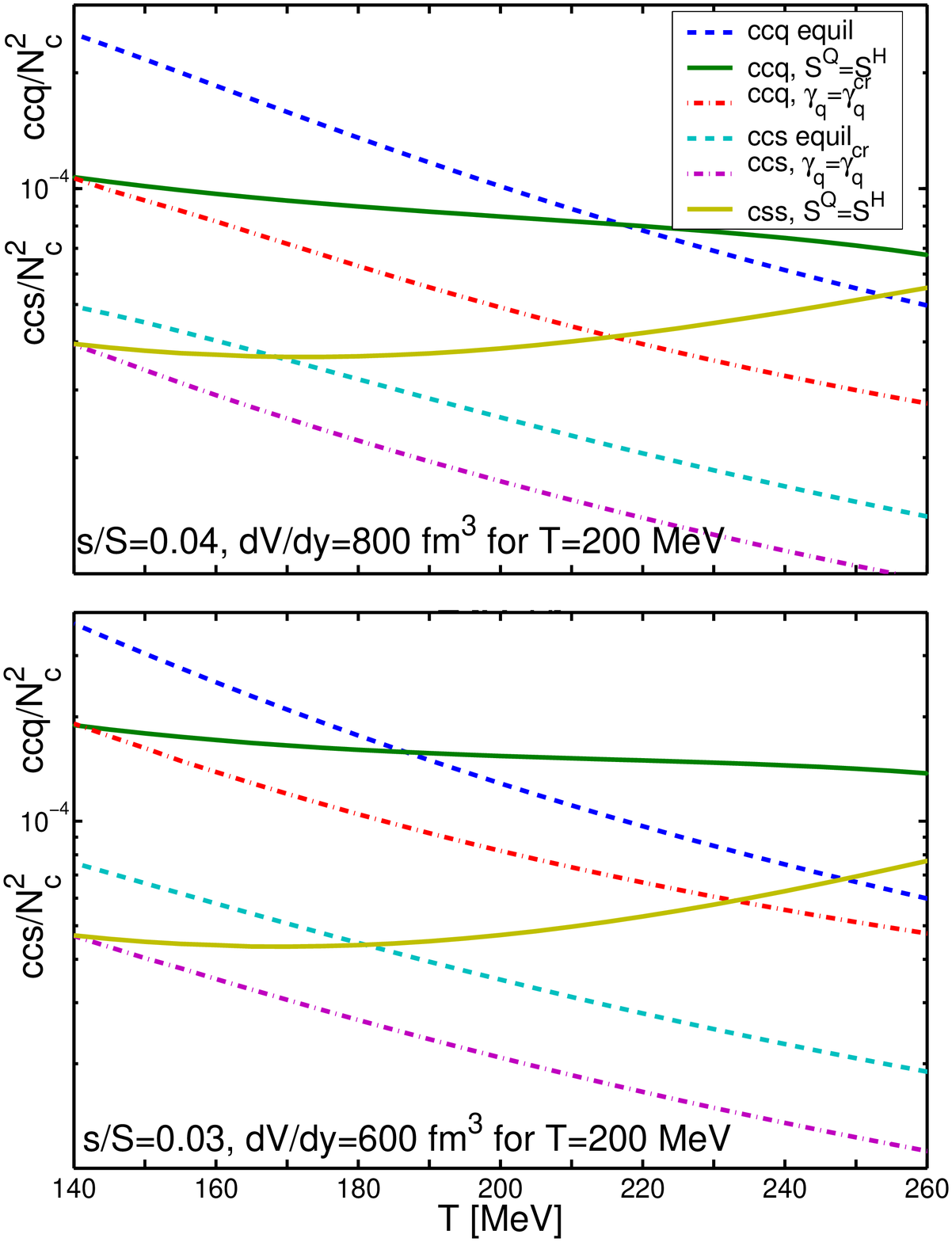}
\caption{(Color on line) \small{$ccq/N_c^2$ (upper lines in each panel) and $ccs/N_c^2$ (lower lines
in each panel) baryon  yields as a function of $T$. Upper panel:
chemical equilibrium case with $dV/dy=800$ $\mathrm{fm^{-3}}$ for $T=200$
MeV(dashed line), $s/S=0.04$ with
$dV/dy=800$ $fm^{-3}$ for $T=200$ MeV: $S^H=S^Q$ (solid
line) and $\gamma_q=\gamma_q^{cr}$ (dash-dot line); and lower panel: chemical equilibrium case with
$dV/dy=600$ $\mathrm{fm^{-3}}$ for $T=200$ MeV (dashed line), and $s/S=0.03$
$S^Q=S^H$ (solid line) and $\gamma^q=\gamma_q^{cr}$ (dash-dot line).}}
\label{qcc}
\end{figure}

In figure~\ref{qcc} we show $N_c^2$ scaled yields of $ccq$ and $ccs$ baryons
as a function of temperature. Upper panel shows aside of the equilibrium case
(dashed lines) the yields for  $s/S=0.04$  with $S^H=S^Q$ (solid lines) and with
 $\gamma_q=\gamma_q^{cr}$ (dash-dot line) for
$dV/dy = 800\,\mathrm{fm^{-3}}$ for $T=200$ MeV. Lower panel is for
$dV/dy = 600\,\mathrm{fm^{-3}}$ and  $s/S=0.03$. For the $ccq$
baryons the chemical non-equilibrium suppression effect is similar
to what we saw for $c\bar c$ and Bc mesons. Equilibrium yield is
much larger than non-equilibrium for $T<230$ MeV when $s/S=0.04$ and
$S^H=S^Q$, and for $T<190$ MeV when $s/S=0.03$ and $S^H=S^Q$. In
case $\gamma_q=\gamma_q^{cr}$, the yield of $ccq$ is always smaller
than equilibrium. The yield of $ccs$ baryons has similar
suppression, but it becomes larger than equilibrium for smaller
temperatures and yield enhancement for higher T is
 larger for $S^H=S^Q$ then in case of $ccq$ because of large number of strange quarks.

\begin{figure}
\centering
\includegraphics[width=8.1cm,height=10.8cm]{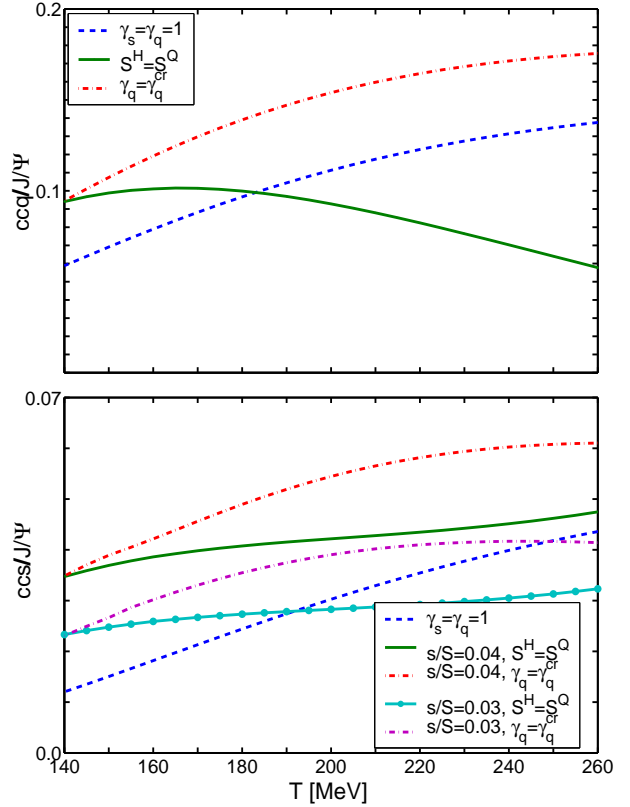}
\caption{(Color on line) \small{$ccq/J\!/\!\Psi$ (upper panel) and $ccs/J\!/\!\Psi$ (lower  panel) ratios as a function of $T$.
Upper panel:
chemical equilibrium case  (dashed line), $S^H=S^Q$ (solid
line) and $\gamma_q=\gamma_q^{cr}$ (dashed-dot line); and lower panel: chemical equilibrium case with
(dashed line), $s/S=0.04$: $S^Q=S^H$ (solid line with dot marker) and $\gamma_q=\gamma_q^{cr}$ (thin dash-dot line); $s/S=0.03$
(solid line) and $\gamma_q=\gamma_q^{cr}$ (thin dash-dot line).}}
\label{qccjpsi}
\end{figure}

 In the figure~\ref{qccjpsi} we show ratios $ccq/J\!/\!\Psi$ (upper panel) and $ccs/J\!/\!\Psi$ (lower panel)
as a function of hadronization temperature. These ratios do not
depend on $dV/dy$.  $ccq/J\!/\!\Psi\propto \gamma_q$ does not depend
on $s/S$. For $ccq/J\!/\!\Psi$ ratio we show three cases: chemical
equilibrium $\gamma_s=\gamma_q=1$ (dashed line), $S^H=S^Q$ (solid
line) and $\gamma_q=\gamma_q^{cr}$ (dash-dot line). For
$ccs/J\!/\!\Psi$ ($ccq/J\!/\!\Psi \propto \gamma_s$) we show
chemical equilibrium case (dashed line), $s/S=0.04$: $S^H=S^Q$
(solid line with point marker) and $\gamma_q=\gamma_q^{cr}$ (thin
dash-dot line); $s/S=0.03$: $S^H=S^Q$ (solid line) and
$\gamma_q=\gamma_q^{cr}$ (thin dash-dot line). The overall all
yields of double charmed (strange and non-strange) baryons and
anti-baryons is clearly larger than the yield of $J\!/\!\Psi$.

\section{Conclusions}\label{concSec}

We have considered here in some detail the abundances
of heavy flavor hadrons within the statistical hadronization model.
While we compare the yields to the expectations based on
chemical equilibrium yields of light and strange quark pairs,
we present results based on the hypothesis that
the QGP entropy and QGP flavor yields determine
the values of phase space occupancy $\gamma^\mathrm{H}_i$ $i=q,s,c,b$,
which are of direct interest in study of the heavy hadron yields.

For highest energy heavy ion
collisions the range of  values discussed  in literature is
$1\le \gamma^{\mathrm{H}}_q\le 1.65$ and
$0.7\le \gamma^{\mathrm{H}}_s/\gamma^{\mathrm{H}}_q\le 1.5$. However
$\gamma^{\mathrm{H}}_c$ and $\gamma^{\mathrm{H}}_b$
values which are much larger than unity arise. This is
due to the need to describe the large primary parton based
production, and considering that the   chemical
equilibrium yields   are suppressed by the factor $\exp(-m/T)$.

Our work is based on the grand canonical treatment of phase space.
 This approach is valid for charm hadron production at LHC,  since
the  canonical corrections, as we have discussed, are
 not material.  On the other hand, even at LHC 
the much smaller  yields of  bottom heavy hadrons are
subject to canonical suppression. The value of the parameter $\gamma_b^{H}$
obtained at a   fixed bottom yield  $N_b$, using either  the canonical, or the grand canonical methods,
are different, see e.g.  Eq.\,(15) in \cite{Rafelski:2001bu}.  Namely, to obtain  a given yield $N_b$ 
in canonical approach, a greater value of $\gamma_b^{H}$ is needed 
in order to compensate the canonical suppression effect. 
However, for any individual single-$b$ hadron, 
the  relative yields, i.g. $B/B_s$ do  not depend on $\gamma_b^{H}$
and thus such ratios are not influenced by  canonical suppression.  Moreover, as long as 
the yield of single-$b$ hadrons dominates the total  bottom yield: 
$N_b\simeq B+B_s+\Lambda_b+\ldots$, also the $N_b$ scaled yields of
hadrons comprising one $b$-quark i.e. ratios such as $B/N_b$, $B_s/N_b$, $B_c/N_b$, etc,  
are not sensitive to the value of   $\gamma_b^{H}$ and  can be obtained
within either the canonical, or grand canonical method. 
 On the other hand for $b\bar{b}$ mesons and multi-$b$ baryons the
canonical effects should be considered. Study of the
yields of these particles is thus postponed.

We address here  in particular how the yields of heavy hadrons are influenced by
$\gamma^{\mathrm{H}}_s/\gamma^{\mathrm{H}}_q\ne 1$ and $\gamma_q \ne 1$. The actual values
of $\gamma^{\mathrm{H}}_s/\gamma^{\mathrm{H}}_q$ we use are related to
the strangeness per entropy yield $s/S$ established in the QGP phase.
Because the final value $s/S$ is established well before hadronization,
and the properties of the hadron phase space are well understood,
the resulting $\gamma^{\mathrm{H}}_s/\gamma^{\mathrm{H}}_q$ are well
defined and turn out to be quite different from unity in the range of
temperatures in which we expect particle freeze-out to occur.
We consider in some detail the effect of QGP hadronization on
the values of $\gamma^{\mathrm{H}}_s$ and $\gamma^{\mathrm{H}}_q$.

One of first results we present (figure~\ref{JpsiD}) allows a test of the
statistical hadronization model for heavy flavor:
 we show that the yield ratio
$c\bar c$ $s\bar s$/($c\bar s$ $\bar c s$) is nearly independent
of temperature and it is also nearly constant when the $\phi$ is
allowed to freeze-out later (figure~\ref{JpsiD2T}), provided that the condition of
production is at the same value of strangeness per entropy $s/S$.

We studied in depth how the (relative) yields of strange and non-strange
charmed mesons vary with strangeness content. For a chemically
equilibrated QGP source, there is considerable shift of the yield
from non-strange $D$ to the strange $D_s$
 for $s/S=0.04$ expected at LHC.
The expected fractional yield $D_s/N_c \simeq {B_s}/N_b\simeq 0.2$
when one assumes $\gamma^{\mathrm{H}}_s=\gamma^{\mathrm{H}}_q=1$,
 the expected
enhancement of the strange heavy mesons is at the level of 30\%
when $s/S=0.04$, and greater when greater strangeness yield is
available.

A consequence of this result is that we find a relative suppression of the
multi-heavy hadrons, except when they contain strangeness. The somewhat
ironic situation is that while higher beyond chemical equilibrium charm QGP yield enhances production
of $c\bar c$ states,  beyond chemical equilibrium enhanced light quarks and strangeness
multiplicity suppresses this almost by
that much. This new phenomenon adds to the complexity of interpretation
of hidden charm meson yield. On the other hand, the yield of
$c\bar{c}/N_c^2 \simeq 2 10^{-3}$ is found to be
almost independent on  hadronization
temperature in case  which conserves entropy at hadronization.
We don't know exactly equation of state in QGP and so the value of
 $\gamma_q$ which is needed to conserve the entropy may be different.
 If $\gamma_q$ is larger for higher temperatures, suppression of
$c{\bar{c}}$ is larger for a fixed $s/S$.
The same result is found for $B_c \approx 5-6\,10^{-4} N_cN_b $,
that  yield remains considerably larger (by a factor 10 --- 100) compared to the scaled
yield in single nucleon nucleon collisions.

We have shown that the study of heavy flavor hadrons will provide
important information about the nature and properties of the QGP
hadronization. The yield of Bc($b\bar c$) mesons remains enhanced
while the hidden charm $c\bar c$ states encounter another suppression
mechanism, compensating for the greatly enhanced production due to
large charm yield at LHC.

\vspace*{.2cm}
\subsubsection*{Acknowledgments}
Work supported by a grant from: the U.S. Department of Energy  DE-FG02-04ER4131.




\vspace*{-0.3cm}

\end{document}